\documentclass[preprint]{aastex}
\usepackage{epsf}
\usepackage{emulateapj5}
\usepackage{onecolfloat}

\def\lsim{\lower0.6ex\vbox{\hbox{$ \buildrel{\textstyle <}\over{\sim}\ $}}}
\def\gsim{\lower0.6ex\vbox{\hbox{$ \buildrel{\textstyle >}\over{\sim}\ $}}}

\def\beq{\begin{equation}}
\def\eeq{\end{equation}}

\def\Om{\Omega_{\rm M}}
\def\Ol{\Omega_{\rm \Lambda}}

\def\Ow{\Omega_{\rm W}}

\def\lcdm{$\Lambda$CDM}

\def\himpc{{h$^{-1}$ Mpc}~}
\def\kms{\textrm{ km s}^{-1}}

\def\Owdm{\Omega_{\rm W}}
\def\dd{\textrm{d}}

\def\cf{{\it cf.}~}
\def\eg{{\it e.g.},~}
\def\ie{{\it i.e.},~}

\def\3he{$^3$He}
\def\4he{$^4$He}

\def\6li{$^6$Li}
\def\7li{$^7$Li}
\def\8Be{$^8$Be}
\def\9B{$^9$B}

\def\Msun{M$_{\odot}$~}

\def\Mvir{M_{\rm vir}}
\def\Rvir{R_{\rm vir}}
\def\cvir{c_{\rm vir}}
\def\Vvir{V_{\rm vir}}
\def\Vmax{V_{\rm max}}
\def\dvir{\Delta_{\rm vir}}

\def\dv2{\Delta_{\rm V/2}}
\def\Rv2{r_{\rm V/2}}

\def\sig8{\sigma_8}

\def\Msat{M_{\rm sat}}

\def\Mhost{M_{\rm host}}

\def\t0{t_0}
\def\lnL{\ln(\Lambda)}

\def\Vorb{V_{\rm orb}}
\def\rt{r_{\rm t}}
\def\Rcirc{R_{\rm circ}}

\def\rmaxsat{r_{\rm max}^{\rm sat}}
\def\rssat{r_{\rm s}^{\rm sat}}

\def\Rmax{r_{\rm max}}

\def\svd{\sigma_{\star}}

\def\ipl4{IPL4}

\def\lb{\left [}
\def\rb{\right ]}

\begin{document}


\slugcomment{{\em The Astrophysical Journal, submitted}}

\twocolumn[
\lefthead{Halo Substructure and the Power Spectrum}
\righthead{Andrew R. Zentner \& James S. Bullock}

\title{Halo Substructure and the Power Spectrum}\vspace{3mm}
%

\author{Andrew R. Zentner}
\affil{Department of Physics, The Ohio State University, 174 W.
18th Ave, Columbus, OH 43210-1173}
\email{zentner@pacific.mps.ohio-state.edu}
\author{James S. Bullock\altaffilmark{1}}
\affil{Harvard-Smithsonian Center for Astrophysics, 60
Garden St., Cambridge, MA 02138}
\email{jbullock@cfa.harvard.edu}
\altaffiltext{1}{Hubble Fellow}
\footnotetext{hi}
%
%
\begin{abstract}

We  present a semi-analytic model  to  investigate the merger history,
destruction rate,   and    survival probability of    substructure  in
hierarchically  formed dark  matter  halos, and use  it  to  study the
substructure content of halos  as a function  of input primordial power spectrum.
For    a  standard    cold   dark    matter  ``concordance'' cosmology
($\Lambda$CDM; $n=1$, $\sigma_8 = 0.95$) we successfully reproduce the
subhalo velocity  function and  radial  distribution profile  seen  in
N-body   simulations, and  determine   that the  rate   of merging and
disruption  peaks  $\sim   10-12$  Gyr  in the   past for 
Milky Way-like halos, while surviving
substructures are typically  accreted within the  last $\sim 0-8$ Gyr.
We  explore power spectra   with  normalizations  and spectral
``tilts'' spanning the ranges $\sigma_8 \simeq 1 - 0.65$ and $n \simeq
1-0.8$, and include a  ``running-index'' model with  ${\rm d} n / {\rm
d} \ln   k = -0.03$ similar  to  the best-fit  model discussed  in the
first-year  WMAP  report.    We investigate  spectra   with  truncated
small-scale power, including a  broken-scale inflation model and three
warm dark matter cases with $m_{\rm W} = 0.75 - 3.0$ keV.

We   find  that  the  mass   fraction  in  substructure is  relatively
insensitive  to the tilt and   overall normalization of the primordial
power   spectrum.  All of    the   CDM-type  models yield    projected
substructure mass fractions that  are consistent with,
but on the low side, of 
published  estimates from strong lens systems: $f_{9} =  0.4 -
1.5\%$ ($64$ percentile), for  subhalos  $< 10^{9}$ \Msun  within
projected cylinders of radius $r < 10$  kpc.  Truncated models 
produce
significantly smaller fractions, $f_{9} = 0.02 - 0.2\%$ for
$m_{\rm W}\simeq 1$ keV, and are disfavored by lensing
estimates of substructure. 
This suggests  that  lensing and
similar probes can provide a
robust test  of   the CDM  paradigm   and  a powerful   constraint  on
broken-scale  inflation/warm particle masses,  including masses larger
than the $\sim 1$ keV upper limits of previous studies.
We  compare our  predicted  subhalo velocity  functions  to the  dwarf
satellite  population of the Milky  Way.  Assuming dwarfs have
isotropic velocity dispersions, we find that  the standard $n=1$ model
over-predicts the  number of Milky  Way satellites  at $\Vmax \lsim 35
\kms$, as expected.    Models with less  small-scale power   do better
because  subhalos  are  less   concentrated and   the mapping  between
observed   velocity   dispersion  and  halo  $\Vmax$  is significantly
altered.  The running-index model, or a fixed tilt with $\sigma_8
\sim 0.75$, can account for the local dwarfs without the need  for  
differential feedback   (for  $\Vmax \gsim  20 \kms$); however, 
these comparisons depend sensitively  on  the assumption of
isotropic velocities in satellite galaxies.

\end{abstract}
%
%
%

]
%
%
%
\section{\label{sec:intro}INTRODUCTION}

\footnotetext{Hubble Fellow}

In the standard cosmological model of structure formation ($\Lambda$CDM), the
Universe is  dominated by cold, collisionless dark  matter (CDM), made
flat by a cosmological  constant ($\Lambda$), and endowed with initial
density perturbations  via quantum fluctuations  during inflation. The
\lcdm~ model  with $\Om = 1-\Ol  = 0.3$, $h \approx  0.7$, and a
scale-invariant  spectrum of  primordial perturbations  ($P(k) \propto
k^n$,  $n=1$,  $\sigma_8  \sim   0.9$)  is  remarkably  
successful  at reproducing a plethora of large-scale observations 
(\eg Spergel et al. 2003; Percival et al. 2002).   In
contrast,  several small-scale observations  have proven  more  difficult to
explain.   Galaxy densities and concentrations appear 
to  be much  lower than what  is predicted for the  standard ($n=1$) 
$\Lambda$CDM model (\eg Debattista \& Sellwood 2000; C{\^o}te, Carignan, \& 
Freeman 2000; Borriello \& Salucci 2000; Binney \& Evans 2001; 
Keeton 2001; van den Bosch \& Swaters 2001; Marchesini et al. 2002; 
Swaters et al. 2003; McGaugh, Barker, \& de Blok 2003; 
van den Bosch, Mo, \& Yang 2003), 
and the Local Group dwarf  galaxy  count is  significantly  below  what 
might naively  be expected from the substructure content of $\Lambda$CDM  
halos (Klypin et al. 1999a, K99 hereafter; Moore et al. 1999a).  
In Zentner \& Bullock (2002, hereafter ZB02), we showed that the central densities of
$\Lambda$CDM dark matter  halos can be brought into  reasonable agreement
with  the rotation curves of  dark matter-dominated galaxies by 
reducing galactic-scale  fluctuations  in the  initial  power spectrum
($\sigma_8 \sim  0.75$ and $n  \sim 0.9$ is a good match;  see
Alam, Bullock, \& Weinberg 2002, hereafter  ABW; McGaugh et al.  2003;
van den Bosch et al. 2003).  The present paper is an  extension of this work.
We  explore how changes   in the  initial   power spectrum affect  the
substructure    content of $\Lambda$CDM halos, test   our findings against
attempts to measure the  substructure mass fraction  via gravitational
lensing, and relate our results to the question of the abundance of 
dwarf satellites in the Local Group.

It is straightforward to see  why CDM halos are  expected to play host
to a large number of  distinct, bound substructures, or  ``subhalos.''
In the  modern picture of hierarchical structure  formation (White \& Rees 1978;
Blumenthal et al.  1984; Kauffmann, White, \& Guiderdoni 1993) low-mass systems collapse
early and merge to form larger systems over time.  Small
halos collapse at high  redshift, when the universe  is very dense, so
their central densities  are  correspondingly high.  When these  halos
merge into larger hosts, their high densities allow them to resist the
strong tidal  forces that  act to  destroy them.   While gravitational
interactions do serve to unbind  most of mass associated with merged progenitors, 
a significant fraction of these small halos survive as distinct substructure.

Our  understanding of this process   has increased dramatically in the
last five years thanks to remarkable advances in   N-body techniques
that allow the  high force and mass resolution necessary to study 
substructure in detail (Ghigna et al. 1998, 2000; Kravtsov 1999; K99; 
Klypin et al. 1999b; Kolatt et al. 1999; Moore et al. 1999a,b; Font et al.
2001; Stoehr et  al.  2002).  For  $n=1$, $\Lambda$CDM and CDM simulations,
the total mass  fraction  bound up in  substructure  is measured at  $f \sim
5-15\%$ (Ghigna et al.  1998, Klypin et al. 1999b), with a significant
portion contributed by  the  most massive subsystems, $dN/dM   \propto
M^{-s}$, $s \approx 1.7$.  The  substructure content of halos seems to
be roughly  self-similar when subhalo mass is scaled
by the host halo mass (Moore et al. 1999a) and the subhalo count is observed to decline at 
the  host halo center, where tidal forces  are strongest 
(Ghigna et al. 1998; Col{\'{\i}}n et al. 2000b; Chen, Kravtsov, \& Keeton 2003).

Unfortunately, studies of substructure using N-body simulations
suffer from issues of numerical resolution.  Simulations with the capability 
to resolve substructure are computationally expensive.  
They cannot be used to study the implications of many unknown
input parameters and cannot attain both the resolution and the 
statistics needed to confront observational data on substructure that 
appear to be on the horizon.  Even state of the art simulations 
face difficulties in  the  centers of halos where ``overmerging'' may be 
a problem (\eg Chen et  al.  2003; Klypin et al. 1999b) and 
measurements of the substructure fraction  via lensing are highly sensitive to these
uncertain, central regions.   Our goal is  to  present and apply a
semi-analytic model that suffers from  no inherent resolution effects,
and is based on the processes that were observed to govern 
substructure populations in past N-body simulations.   
This kind  of  model can  generate statistically
significant predictions for  a variety of inputs quickly 
and can  be used  to  guide   expectations for the  next
generation of N-body simulations.  Conversely, this model represents 
in many ways an  extrapolation of N-body results into unexplored 
domains and it is imperative that our results be tested by future numerical 
studies.  In the present paper, we aim to explore the effect of the 
power spectrum on the population of
surviving subhalos, but in  principle these methods  are  suitable for
testing substructure ramifications for a variety of cosmological inputs.

One of the main  motivations  for this work comes
from simulation results that indicate that galaxy-sized CDM halos 
play host to hundreds of subhalos with maximum circular
velocities in the range $10 \kms$ $\lsim$ $\Vmax$ $\lsim$ $30$ $\kms$.
The Milky  Way,  as  a comparative example, hosts only $11$   dwarf
satellites of similar size.  This ``dwarf satellite problem'' 
specifically refers to the gross  mismatch between the predicted
number of $\Lambda$CDM subhalos and the count of satellite galaxies in the
Local  Group (K99; Moore et al. 1999a; Font et al. 2001; 
see also Kauffmann et al. 1993, who indicated
that  there may  be a problem  using  analytic arguments).  The  dwarf
satellite problem  and other  small-scale issues led  many  authors to  consider
modifications  to  the standard framework.    If  the dark matter were
``warm'' (Pagels \& Primack 1982; Colombi, Dodelson, \& Widrow 1996; Hogan
\& Dalcanton  2000; Col{\'{\i}}n et al. 2000a; Bode, Ostriker, \& 
Turok 2001; Lin et al. 2001; Knebe  et al. 2002) or if  
the  primordial power spectrum  were sharply truncated on small scales
(Starobinsky 1992; Kamionkowski \& Liddle 2000) then subgalactic-scale
problems may be  allayed  without  vitiating  the overall success   of
$\Lambda$CDM on large scales. 
Another possibility is that CDM substructure is abundant in all galaxy
halos, but  that  most low-mass  systems  are simply  devoid of
stars.  An intermediate solution  may involve  a
simple modification  of   the assumed primordial  spectrum  of density
perturbations that gradually lowers power 
on  galactic scales  relative to the horizon, \eg 
via tilting the power spectrum.

Probing models with low galactic-scale power  is motivated not only by
the small-scale crises facing standard $\Lambda$CDM but also by more direct
probes of the power spectrum.  While many analyses continue to measure
``high'' values for $\sig8 \sim 1$  (Van Waerbeke et al. 2002; Komatsu
\& Seljak 2002; Bahcall \& Bode 2003; where $\sig8$ is the linear, rms
fluctuation amplitude on a length scale of  8 \himpc), numerous recent
studies relying on similar techniques, advocate rather ``low'' values
of $\sig8  \sim 0.7-0.8$ (Jarvis et al. 2003; Bahcall et al. 2003; 
Schuecker et al 2003; Pierpaoli et al. 2003; Viana et al. 2002; Brown et al. 2002; 
Allen  et al. 2002; Hamana  et al.  2002; Melchiorri \& Silk 2002; Borgani et al. 2001).  
Similarly, the Ly-$\alpha$ forest measurements of the power  spectrum
are  consistent  with reduced galactic-scale power (Croft et al. 1998; 
McDonald et al.  2000; Croft et al.  2002).  Set against the
normalization of fluctuations on   large scales implied by  the Cosmic
Background Explorer (COBE) measurements of cosmic microwave background
(CMB)  anisotropy (Bennett et al. 1994), these data suggest that 
the initial power spectrum may be tilted to favor large scales with $n <1$.  

The recent analysis of the Wilkinson Microwave Anisotropy Probe (WMAP)
measurements of CMB anisotropy presented by Spergel  et al. (2003; see
also    Verde et al.   2003; Peiris  et al.  2003)  returns a best-fit
spectral index to a pure power law primordial  spectrum of $n=0.99 \pm
0.04$ when only the WMAP data are considered.  However, when data from
smaller scale  CMB experiments, the  2dF Galaxy Redshift  Survey, and
the Ly-$\alpha$ forest are included, the analysis  favors a mild tilt,
$n = 0.96  \pm 0.02$.  Interestingly, all  of  the data  sets together
yield  a better fit if the  index is allowed to run:  the WMAP 
team find ${\rm d}n/\rm{d}\ln k =  -0.031^{+0.016}_{-0.017}$.
This result is consistent with no  running at $\sim 2\sigma$,
and the statistical significance is further weakened 
when additional uncertainties in the  mean flux decrement in the 
Ly-$\alpha$  forest  are considered
(Seljak, McDonald, and Makarov 2003; Croft et al. 2002), yet such
a model certainly seems  worth investigating, especially 
in light of the small-scale difficulties it may help to alleviate.

\begin{figure}[t]
\begin{center}
\resizebox{!}{9cm}
{\includegraphics{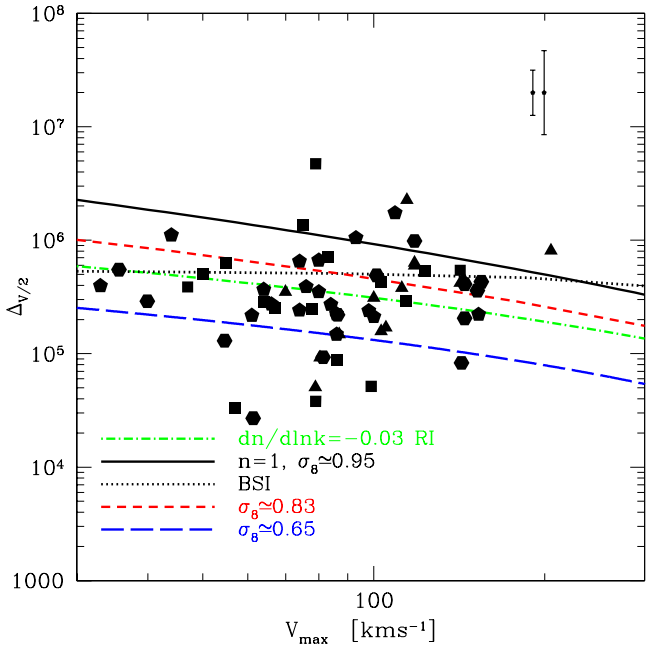}}
\caption{\label{fig:dv2} {Galaxy central densities.  The symbols show
the mean density (relative to the critical  density) within the radius
where each rotation curve falls to half of its maximum value, inferred
from the  measured  rotation curves  of several  dark matter-dominated
dwarf and low surface brightness galaxies (see ZB02 for details).  The
data  are taken from de Blok,  McGaugh, \&  Rubin (2001; triangles and
hexagons),  de Blok and  Bosma   (2002; squares), and  Swaters  (1999;
pentagons).  The lines show the theoretical expectation for several of
the power spectra we describe in \S \ref{sec:ps} and Table
\ref{table:spectra}.  The points with error bars in the upper right corner 
reflect the expected theoretical scatter in the density as inferred by 
Bullock et al. (2001, larger range) and Jing (2000, smaller range).
}}
\end{center}
\end{figure}

We  explicitly show how   models  with reduced  small-scale power  are
expected  to   help  the halo    density    problem  in Figure
\ref{fig:dv2}, which is an updated version of Figure 5 in ZB02.  Here, we 
compare the densities of standard $n=1$  halos to
galaxy rotation curve data (see ZB02 for details) along
with   expectations for the running-index (RI) model favored by WMAP 
and several other models we explore in the following sections.  
Galaxy and halo  densities  (vertical axis) are evaluated  at
the radius where  the rotation curve  falls to half its  maximum value
and expressed in units  of  the critical density ($\Delta_{V/2}$,   as
defined in  ABW).   Clearly, the  data   favor low small-scale   power
relative  to the standard  $n=1$  case.

The possibility of discriminating between  standard $\Lambda$CDM and 
several alternatives has inspired efforts to
measure  and quantify the substructure  content of galactic halos.  One 
method relies on   studying tidal tails  associated
with  known  Galactic satellites  (Johnston,  Spergel, \&  Haydn 2002;
Ibata et al. 2002a, 2002b; Mayer et  al. 2002).  
Subhalos passing through cold
tidal  streams  scatter  stars away  from their original
orbits,  and the signatures of  these events may  be detectable in the
velocity data of future astrometric missions and several deep
halo surveys that will soon  be completed.  Of particular interest for
obtaining measurements in distant galaxies are studies that aim to
detect substructure  via flux ratio anomalies in strong gravitational lenses
(Moore et al. 1999a; Metcalf \& Madau 2001; Metcalf \& Zhao 2002; 
Brada{\v c} et al. 2002).  Using a sample of  seven four-image radio lenses, 
Dalal \& Kochanek (2001, DK01 hereafter) estimated a mass  fraction of 
$f = 0.006  - 0.07$ (90\% confidence level) 
bound up in substructure less massive than $\sim  10^8 -  10^{10}$ M$_{\odot}$, 
in line   with the rough  expectations of CDM.\footnote{DK01  
quoted an approximate upper mass limit of  $10^6 - 10^{9}$ M$_{\odot}$.
They have since concluded that an upper limit of $\sim 10^8 - 10^{10}$
M$_{\odot}$ may be more appropriate (N. Dalal, private communication).}
While measurements   of   this kind   are  susceptible  to   potential
degeneracies   with  the  adopted    smooth   lens model   and   other
uncertainties, they are encouraging  and serve as prime motivators for
this work (see Kochanek \& Dalal 2003).  In addition, new
observational techniques that focus  on astrometric features  (Metcalf
2002), and particularly spectroscopic   studies of   strong  lens
systems (Moustakas \& Metcalf 2003), promise to hone  in on the masses
of the subclumps responsible for the lensing signals.

If the Milky  Way really is surrounded  by a large  number of
dark subhalos, the dwarf satellite problem serves  as a conspicuous
reminder  that feedback plays an  important role  in hierarchical
galaxy formation.  Of  course, the need for feedback  in small systems
has been generally recognized for  some time  (\eg White  \& Rees  1978).  
Supernova  blow-out likely plays a role in regulating
star formation if CDM is the  correct theory (Dekel \& Silk 1986;
Kauffmann et al. 1993;  Cole et al.  1994; Somerville \&
Primack  1999); however, supernova  winds do not naturally suggest a
sharp discrepancy at $\sim 30 \kms$, nor do they explain why some halos 
of this size should have stars while  most have none at all.  It seems 
more likely that supernovae play an important role in setting scaling 
relations in slightly larger galaxies ($\Vmax \sim 100 \kms$; 
Dekel \& Woo 2002; but see Mac Low \& Ferrara 1999).  
Perhaps a more natural feedback source on satellite galaxy scales 
is the ionizing background, which should suppress  galaxy formation 
in halos with  $\Vmax \lsim 30 \kms$  (Rees 1986;  Efstathiou  1992;  
Kauffmann  et al. 1993; Shapiro, Giroux, \&  Babul 1994; Thoul \& Weinberg 
1996; Quinn, Katz, \&  Efstathiou 1996; Gnedin 2000).  
Bullock,  Kravtsov, \&  Weinberg (2000, BKW  hereafter) suggested that 
dwarf galaxies may be associated with small halos that collapsed before the
epoch of reionization, and though  the method used
by BKW to estimate dwarf luminosities was crude,  more sophisticated
models have since led to similar conclusions  (Chiu, Gnedin, \& Ostriker 2001; 
Somerville 2002; Benson  et  al. 2002).  For the smallest
systems, $\Vmax \lsim 10-20 \kms$, the ionizing background likely stops
star formation altogether by  photo-evaporating gas in halos, even
after they have collapsed (Barkana \& Loeb 1999; Shaviv \& Dekel 2003).

Precisely what can be  learned about galaxy formation and/or cosmology
by counting dwarf  satellites depends upon our expectations
for the density profiles of their host halos.  To count
satellites of a given maximum circular velocity, we  must infer a halo
$\Vmax$    using   the     observed     central  velocity   dispersion
$\sigma_{\star}$, and the mapping between these two quantities depends
sensitively on the structure of each satellite's  dark  matter
halo.  This point was first emphasized by S. D. M. White at the Summer
$2000$ Institute for Theoretical  Physics Conference on Galaxy Formation
and Evolution.\footnote{See http://online.kitp.ucsb.edu/online/galaxy\_c00/white/.}
Standard $\Lambda$CDM   halos with $\Vmax \sim  30 \kms$ are expected  to be
very concentrated (Col{\'{\i}}n et al. 2000a; 
Bullock et al. 2001), with their rotation curves peaking at $r_{\rm max} \sim
1$ kpc, so the multiplicative factor that converts a $\sim 1$ kpc 
velocity dispersion measurement to halo $\Vmax$ is fairly modest:  $\Vmax
\sim \sqrt{3} \sigma_{\star}$ (K99).  
However, as we discuss in \S 4,  the  appropriate  conversion  is
cosmology-dependent because models with later structure formation tend
to produce halos  with more slowly rising rotation  curves.
If a dwarf galaxy sits in a halo with a slowly rising rotation curve 
that peaks at $r_{\rm max} \gg$ a few kpc, the conversion factor, and thus 
$\Vmax$, can be significantly larger.  
Shifts   of  this  kind  in  the ``observed'' velocity function 
change the implied velocity (or mass) scale of discrepancy, 
and influence our ideas about the type of feedback that gives rise 
to  the mismatch.  

Hayashi   et al. (2003, H03 hereafter)  and Stoehr et al. (2002, S02
hereafter) suggested   that  substructure  halos   experience
significant mass redistribution in their centers  as a result of tidal
interactions and that they are  therefore less concentrated than comparable 
halos in the field.  They  argue that when  this is taken into  account, the
dwarf satellite mismatch sets in at $\Vmax
\sim 20 \kms$, and that the transition is sudden --- 
below  this scale  all  halos  are devoid of   observable
galaxies.  While these  conclusions have yet  to be confirmed and  are
dependent  upon  subhalo merger  histories  and the isotropy  of dwarf
velocity   dispersions, they    highlight the  need   to   refine  our
predictions about halo substructure.  They also motivate us to explore
how   minor changes   in   cosmological parameters  can influence  our
interpretation of the dwarf satellite problem.

In the remainder of this paper we present our study of CDM substructure. 
In \S\ref{sec:model}, we describe our semi-analytic model, provide
some illustrative examples,  and 
compare our results for standard $\Lambda$CDM  to previous  N-body results.
In \S\ref{sec:ps}, we briefly describe the input power spectra
that serve as the basis for this study.
In \S\ref{sec:results}, we present our results on subhalo mass
functions and velocity functions.  
We  make predictions aimed at measuring substructure 
mass fractions via gravitational lensing 
and address the dwarf satellite problem in light of
some of our findings in this section.  
In \S\ref{sec:caveats} we discuss some shortcomings of our 
model and how they might be improved in future work.
In  \S\ref{sec:conclusions} we summarize 
our work and draw conclusions from our results.  In this 
study we vary only the power spectrum and work within the context of
the so-called ``concordance'' cosmological  model with $\Omega_{\rm
M}=0.3$,  $\Omega_{\rm  \Lambda}=0.7$,   $h=0.72$,  and   $\Omega_{\rm
B}h^2=0.02$ (\eg Turner 2002; Spergel et al. 2003).
%
%
%
%
%
%
%
%
%

\section{\label{sec:model}MODELING HALO SUBSTRUCTURE}
%
%

In order to  determine the substructure   properties of a  dark matter
halo we must  model its mass accretion history  as well as the orbital
evolution of  the  subsystems once they  are  accreted.  For  the first
step,  we rely on the the extended Press-Schechter (EPS)
formalism to create merger histories for  each host system.  We give a 
brief description of our EPS merger trees in \S
\ref{sub:model}.  In   \S \ref{sub:ppt} we   discuss our model for the
density structure of accreted halos and the host system and in \S
\ref{sub:desc} we describe our method for following the orbital evolution 
of each merged system.  We show tests and examples of this model in
\S \ref{sub:test}.

\subsection{\label{sub:model}Merger Histories}

We track diffuse mass accretion and satellite halo acquisition of host
systems by constructing merger histories using the EPS method 
(Bond et al. 1991; Lacey  \& Cole 1993,  LC93  hereafter).  In  particular, we
employ the merger tree  algorithm of Somerville  and  Kolatt
(1999,  hereafter SK99).  This   allows us to  generate  a list of the
masses and accretion  redshifts of all subhalos  greater than  a given
threshold mass that  merged to form the  host  halo.  We describe  the
method  briefly here, and encourage  the  interested reader to consult
LC93 and SK99 for further details.

A merger tree that reproduces many
of the  results of N-body simulations can be constructed using
only the linear power spectrum.  For convenience, we express this
in terms of  $\sigma(M)$, the rms fluctuation amplitude  on mass scale
$M$ at $z=0$.  As in LC93, let $S(M) \equiv \sigma^2(M)$, $\Delta
S \equiv S(M) - S(M + \Delta M)$, $w(t) \equiv \delta_{\rm c}(t)$, and
$\delta w \equiv  w(t) - w(t+\Delta t)$.  Here  $\delta_{\rm c}(t)$ is
the  linear overdensity  for collapse  at time  t, associated  with our 
choice of cosmology (see LC93 or White 1996).
The probability that a halo of mass $M$, at time
$t$, accreted an amount of mass associated with a step of $\Delta S$, in
a given time step implied by $\delta w$ is 
\beq
\label{eq:merger}
P(\Delta S, \delta w) \textrm{d}(\Delta S) =
\frac{\delta w}{\sqrt{2\pi}\Delta S^{3/2}} 
\exp \Bigg[\frac{-(\delta w)^2}{2\Delta S} \Bigg] \textrm{d}(\Delta S).
\eeq
Merger histories  are constructed by starting  at a chosen
redshift and  halo mass  and stepping back  in time  with an
appropriate time step.  
If the minimum mass of a progenitor that we wish to track is 
$M_{\rm min}$, then SK99 tell us that each time step must be small 
in order to reproduce the conditional mass functions of EPS
theory: $\delta w \lsim \sqrt{M_{\rm min}(\textrm{d}S(M)/\textrm{d}M)}$.  

We  build merger trees by selecting progenitors at each time step
according to  equation (\ref{eq:merger}) and treating  events   with
$\Delta M < M_{\rm min}$ as diffuse  mass accretion.  At each 
step, we identify  the most massive progenitor with the host halo and
all less massive progenitors with accreted subhalos and we continue
this process until the host mass falls  below $M_{\rm min}$. 
In  practice, we  use a slightly modified version of the SK99 scheme.    
At each stage we demand that the number of progenitor halos in the 
mass range we consider be close to the mean  value.  As discussed in BKW,
this modification considerably improves the agreement between the
analytically predicted progenitor distribution  and the numerically
generated progenitor distribution.  In what follows we set 
$M_{\rm min} = 10^5$ M$_{\odot}$.  Our fiducial, z=0 host mass is 
$1.4 \times 10^{12}$ M$_{\odot}$, but we vary these choices in order
test sensitivity to the host mass and redshift.

\subsection{\label{sub:ppt}Halo Density Structure}
%
%
%
%
Whether a merged system survives or is destroyed depends 
on the density structure of the subhalo and on the
gravitational potential of the host system.  Therefore, it 
is worthwhile to describe our assumptions about CDM 
density profiles in some detail.  The size of a virialized 
dark matter halo can be quantified
in terms of its virial mass $\Mvir$, or equivalently its virial radius
$\Rvir$, or virial velocity $\Vvir^2 \equiv G\Mvir/\Rvir$.  The virial 
radius of a halo is defined as the radius within which the mean density
is equal to the virial overdensity $\dvir$, multiplied by the
mean matter density of the Universe $\rho_{\rm M}$, so that 
$\Mvir = 4 \pi \rho_{\rm M} \dvir(z) \Rvir^3/3$.
The virial overdensity $\Delta_{\rm vir}$, can be estimated using 
the spherical top-hat collapse approximation and is 
generally a function of $\Omega_{\rm M}$, $\Omega_{\rm \Lambda}$, and 
redshift (\eg Eke, Navarro, \& Frenk 1998).  We 
compute $\Delta_{\rm vir}$ using the fitting function of Bryan 
\& Norman (1998).  In the cosmology considered here, 
$\dvir(z=0) \simeq 337$, and at high redshift  
$\dvir \rightarrow 178$, approaching the 
standard cold dark matter (\ie $\Omega_{\rm M} = 1$) value. 

The gross structure of dark matter halos has been described by several analytic
density profiles that have been proposed as good approximations to the
results of high-resolution N-body simulations (Moore et al. 1999b; Power et al. 2003). 
In the interest of simplicity, we choose to model all halos with the density 
profile proposed by Navarro, Frenk, \& White (1997; hereafter NFW):
\beq
\label{eq:NFW}
\rho(r) = \frac{\rho_{\rm s}}{(r/r_{\rm s})(1+r/r_{\rm s})^2}.
\eeq
For the NFW profile, the amount of mass contained within a
radius r, is 
\beq
\label{eq:MofR}
M(<r) = \Mvir \frac{g(x)}{g(\cvir)}
\eeq
where $x \equiv r/r_{\rm s}$, $g(y) \equiv \ln(1+y) - y/(1+y)$, and
the concentration parameter is defined as
$\cvir \equiv \Rvir/r_{\rm s}$.  Restating equation (\ref{eq:MofR})
in terms of a circular velocity profile yields
$V_c^2(r) = \Vvir^2  \cvir g(x)/x g(\cvir)$.  The 
maximum circular velocity occurs at a radius
$r_{\rm max} \simeq 2.16r_{\rm s}$, with a value
$\Vmax^2 \simeq 0.216 \Vvir^2 \cvir/g(\cvir)$.

As a result of the study by Wechsler et al.  (2002; W02 hereafter) and
several  precursors   (\eg Zaroubi and  Hoffman   1993; NFW;  Avila-Reese,
Firmani, \& Hern{\'a}ndez 1998; Bullock  et al.  2001, hereafter B01),
we now understand that  dark matter halo concentrations are determined
almost  exclusively by their  mass    assembly histories.  The   gross
picture  advocated by W02 is  that the rate  at  which a halo accretes
mass determines how close to  the center of the host halo
the accreted mass is deposited.  When the mass accretion rate is high,
near  equal mass mergers are  very likely, and dynamical friction acts
to deposit  mass  deep into the   interior of the   host.  After an early
period  of  rapid  mass accretion, the  central densities of halos
remain constant at  a value proportional  to the  mean density  of the
Universe  at the so-called ``formation   epoch'' $z_{\rm c}$, defined as the
redshift when the relative mass accretion rate was similar to the rate
of universal  expansion (see W02 for details).  For typical halos, this 
formation epoch occurred at a time when halos were roughly $\sim 10-20\%$ 
of their final masses.  Additionally, W02 found that the scale radius and central 
density of the best fit NFW profile remain practically constant after the 
initial phase of rapid accretion.  After this time, the mass increase is slow,
and as the virial radius of the  halo grows, its concentration 
decreases as $\cvir \propto (1+z)^{-1}$.

These results lend support to B01, who explained the
observed trends   with halo  mass and  redshift  using a simple,
semi-analytic model that we adopt in this study.  In the B01 model, halo
concentrations $\cvir(M,z)$, depend only on  the value of $\sigma(M)$
and the evolution of linear perturbations, $\delta (z) / \delta (z=0) $.  Specifically,
the  density of a  halo of  mass  $M$ is  set  by  the density  of the
Universe at the time when systems of mass $\sim 0.01 M$ were typically
collapsing.  The collapse epoch $z_{\rm c}$, is defined by 
$\sigma(0.01 M) \equiv \delta_{\rm c}(z_{\rm c})$.  Again, $\delta_{\rm c}(z)$ 
is the linear overdensity for collapse at redshift $z$.  Central
densities determined in this manner connect well to the
findings  of W02.  Halo density  structure  is set at a
time of  rapid accretion, when  progenitor masses typically  were 
$M_{\rm prog} \sim 0.1 M$.  Most of  the mass in a halo  at any given time is
set by accretion events with subhalos of mass $\sim 1/10$ the host halo mass 
(\cf \S \ref{sec:results}).  Thus the  period of
rapid mass accretion involves objects of mass  $\sim 0.1 M_{\rm prog} \sim
0.01 M$, and it is the collapse times and densities of these constituents 
that set the central density of the mass $M$ halo.  

The B01 model reproduces N-body results for $n=1$, $n=0.9$, and 
power-law CDM  models (\eg Col{\'{\i}}n et al. 2000b; 
B01) as well the  WDM simulations of 
Col{\'{\i}}n et al. (2000a) and Avila-Reese et al. (2001).  
However, we stress that N-body tests were restricted to the mass range 
$\sim 10^{9}-10^{14}$ M$_{\odot}$ because of the limited dynamic range 
of numerical experiments.\footnote{Preliminary results from new simulation data 
show promising agreement with the B01 model all the way down to $M \sim 10^7$ M$_{\odot}$ 
(P. Col{\'{\i}}n, private communication).}  Nevertheless, we use the B01 
model to compute concentrations for halos with masses $\ll 10^9$ M$_{\odot}$.  
Our results for $M \lsim 10^9$ \Msun may  
be regarded as a ``best-guess'' extrapolation of N-body results.

Before proceeding, we mention an alternate prescription for assigning
$\cvir(M,z)$ proposed  by Eke,  Navarro,  \& Steinmetz (2001, ENS hereafter).  
ENS investigated  the power spectrum dependence of the
$\cvir-\Mvir$ relation for several  $\Lambda$CDM and WDM models.  
While the B01 and ENS recipes for $\cvir(M,z)$  matched well for 
$\Lambda$CDM, the B01 model failed to reproduce the mass dependence seen in 
simulations by ENS for WDM halos with masses smaller  than the ``free-streaming'' mass 
(see \S \ref{sec:ps}).  The four WDM halos simulated by ENS  
with masses small enough to be appreciably affected by free-streaming 
all had  $\cvir$ values that were $\sim 2\sigma$ lower than the 
B01 model.  Based on these data, ENS proposed a model in which halo collapse 
time depends not only on the amplitude of the power spectrum $\sigma(M)$, 
but on an effective overdensity amplitude, 
$\sigma_{\rm eff} \equiv -\sigma(M) \dd \ln \sigma(M)/\dd \ln M$.  
This results in a $\cvir(M)$ relation that increases with mass  for masses  
smaller   than  the truncation scale and decreases at larger masses as in 
$\Lambda$CDM.  By defining  an effective overdensity in this way, ENS 
were able to account for the low $\cvir$ values observed in their WDM simulations 
and still reproduce the redshift and mass dependence seen in $\Lambda$CDM 
simulations.  The slope of the $\cvir$-$\Mvir$ of ENS is 
shallower than the slope predicted by the B01 relation, 
therefore the ENS model also leads to less concentrated 
halos at small mass ($M \lsim 10^{10}$ M$_{\odot}$) even for identical 
input power spectra.  This disparity grows larger when tilted 
and/or running spectra are considered, as in this paper.

Unfortunately, the ENS model cannot be applied  in
the WDM cases we explore because in these models $\sigma(M)$ is very flat on 
scales smaller than the free-streaming mass and the ENS model breaks 
down when $\dd \sigma(M) / \dd M$ becomes very small.  In the ENS model,  
WDM halos smaller than $\sim 1\%$ of the free-streaming  mass 
{\em never} collapse because $\sigma_{\rm eff} \ll 1$.  
In addition to this practical problem, the ENS predictions are not supported by
the results of  Avila-Reese et al. (2001) and Col{\'{\i}}n et al. (2000a).  
Using $\sim 25$ halos, Avila-Reese et al. found WDM halo concentrations to be 
roughly constant with mass down to several orders of magnitude below the 
free-streaming scale, in accordance with the B01 model predictions.  
In light of these difficulties and the discordant results of different 
N-body studies, we have not explored the implications of the ENS model in
this work.  This is not an indictment of the ENS model.  Rather, the results 
of ENS highlight the uncertainty in assigning halo concentrations to low-mass 
systems, especially with power spectra that vary rapidly with scale.  Our choice 
of the B01 relation is a matter of pragmatism and represents a {\em conservative} 
choice in that halos are assigned the higher of the two predictions of $\cvir$ at 
small mass.  Lower $\cvir$ values (in line with ENS expectations) would result
in less substructure and larger deviations from the standard $\Lambda$CDM 
model than the predictions in \S \ref{sec:results}.
%
%
%
%

\subsection{\label{sub:desc}Orbital Evolution}

With  the accretion  history of  the host halo in place,  
and with a recipe in hand that fixes the density structure of
host and satellite halos, the next
step is to track the orbital evolution of accreted systems.  This
is necessary in order to account for the effects of dynamical friction
and mass loss due to tidal forces.  These processes cause most of
the  accreted subhalos either  to sink to  the center of the host halo
and become ``centrally merged,'' or to lose most  of their mass and be
``tidally  disrupted'' and   no   longer  identifiable as     distinct
substructure.  We model these effects using an improved version of the
BKW technique, borrowing   heavily from the dynamical  evolution model
proposed by Taylor and Babul (2001, TB01 hereafter; see also Taylor \&
Babul  2002) and the dynamical friction  studies of Hashimoto, Funato,
\&  Makino  (2002,  HFM02  hereafter)  and  Valenzuela \& Klypin 
(2003; and Valenzuela \& Klypin, in preparation).

We denote the mass of an accreted subhalo as
$M_{\rm sat}$, its outer radius as $R_{\rm sat}$, 
and the accretion redshift as $z_{\rm acc}$.  We set the subhalo 
concentration to the median value given by the B01 model for this
mass and redshift.  Although initially set by the 
virial mass and radius of the in-falling subhalo,
$M_{\rm sat}$ and $R_{\rm sat}$ are
 allowed to evolve with time, as described in more
detail below.
We track the orbit of each subhalo in the potential of its host from the
time of accretion $t_{\rm acc}$, until today ($t_0 \simeq 13.5$ Gyr
in this cosmology) or until it is destroyed.   
The mass accretion history also yields the host halo mass at
each time step.  We fix the density profile of the
host at each accretion time using the median B01 expectation 
for a halo of that mass.  As we mentioned earlier, the 
scale radius and central density of the host remain approximately 
constant.

For the  purpose of tracking each subhalo orbit, we assume the
host  potential to  be both  spherically symmetric {\em and static}.
We update the host halo profile using the B01 expectation at 
each accretion event, but hold it fixed while each orbit is integrated.
While the  approximation of a static  host potential for each orbit
is  not ideal, it allows for an extremely simple 
prescription that significantly reduces the
computational expense of our study.  Moreover, this  approximation is
not  bereft of  physical  motivation.  As  we  discussed above,  halos
observed in numerical simulations appear to form dense central regions
early in their evolution after which their scale radii and central densities 
remain roughly fixed with time.\footnote{The exception to this is the case 
of a late-time merger of  halos of comparable mass in which case the central
densities and scale radii of the participating halos may change
considerably (W02).}  Additionally, we have run test examples 
that include an evolving halo potential (set by the results of W02) 
and find that this addition has a negligible effect on 
the statistical properties of substructure that we are concerned with 
here.

\begin{figure}[t]
\begin{center}
\resizebox{!}{9cm}
{\includegraphics{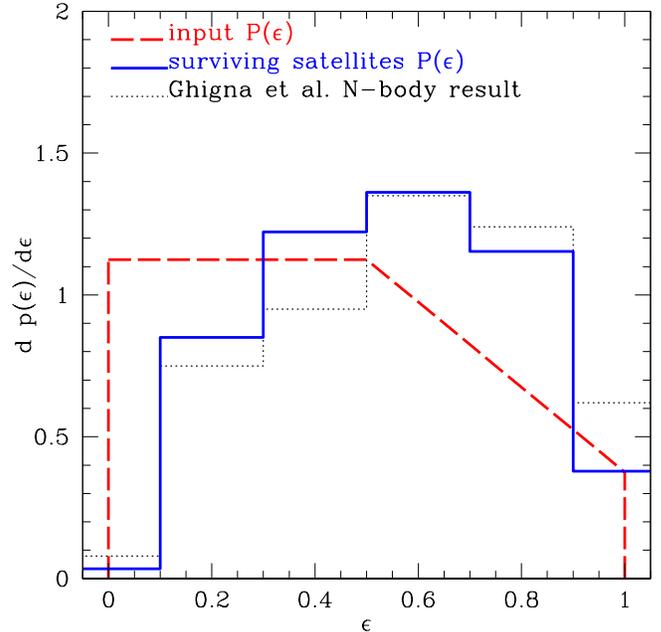}}
\caption{\label{fig:EPS} {Input orbital circularity distribution of
initially in-falling
 substructure (dashed) shown along with the circularity
distribution  of the final surviving population of ($n=1$) LCDM
subhalos at $z=0$ (solid).  For reference, the thin dotted
line shows the circularity distribution of surviving substructure 
measured by Ghigna et al. (1998) in their N-body simulations.}}
\end{center}
\end{figure}

Upon accretion onto the host, each subhalo is assigned an initial 
orbital energy based on the range of binding energies observed in 
numerical simulations (K99; A. V. Kravtsov 2002, 
private communication).  We place each satellite halo on an initial orbit of energy
equal  to the  energy of a circular orbit of radius $\Rcirc  = \eta
\Rvir$, where $\Rvir$ is the virial  radius of the host at the time of
accretion and $\eta$ is drawn randomly from a uniform distribution on
the interval $[0.4,0.75]$.  We assign each satellite an initial
specific angular  momentum $J = \epsilon J_{\rm  circ}$, where $J_{\rm
circ}$ is the specific angular momentum of the aforementioned circular
orbit and $\epsilon$ is  known as the ``orbital circularity.''   Past
studies drew $\epsilon$ from  a uniform  distribution  on the
interval  $[0.1,1]$ (BKW) to match the circularity distribution of {\em surviving} 
subhalos in simulations reported by Ghigna et al. (1998).  However, the
orbits of surviving halos are biased relative  to the orbits  of all
accreted  systems because subhalos on radial orbits are preferentially 
destroyed.  We find that we better match the Ghigna et
al. (1998) result for surviving satellites if we draw the initial  
$\epsilon$ from the simple, piecewise-linear 
distribution depicted in Figure \ref{fig:EPS}.
The initial radial position of each satellite halo is set to $R_{\rm
init} = \Rcirc$ and for all non-circular orbits, we set the subhalo to 
be initially in-falling so that $\textrm{d}R/\textrm{d}t < 0$.
%
%
%
%
%
%

To calculate the trajectories of subhalos, we treat them as point 
masses under the influence of the NFW gravitational potential of
the host halo.   We model orbital decay by dynamical friction using the
Chandrasekhar formula (Chandrasekhar 1943).  The Chandrasekhar 
formula was derived in the context of a highly idealized situation; 
however, numerical studies indicate that this approximate relation 
can be applied more generally (Valenzuela \& Klypin 2003 
have performed a new test that supports the use of this approximation).  
Using the Chandrasekhar approximation, there is a frictional force 
exerted on the subhalo that points opposite to the subhalo 
velocity:
\beq
\label{eq:Fdf}
F_{\rm DF} \simeq \frac{4 \pi \ln(\Lambda) G^2 \Msat^2 \rho(r)}{\Vorb^2}
\Bigg[{\mathrm{erf}}(X) - \frac{2X}{\sqrt{\pi}} \exp(-X^2) \Bigg].
\eeq
In equation (\ref{eq:Fdf}), $\ln(\Lambda)$ is the Coulomb logarithm,
$r$ is the radial position of the orbiting
satellite,  and $\rho(r)$ is the density 
of the host halo at the satellite radius.
The quantity $\Vorb$ is the orbital speed of the satellite halo and 
$X \equiv \Vorb/\sqrt{2\sigma^2}$, where
$\sigma$ is the one-dimensional velocity dispersion of particles in
the host halo.  For an NFW profile, 
the one-dimensional velocity dispersion can be determined
using the Jeans equation.  Assuming an isotropic velocity dispersion tensor, 
\beq
\label{eq:sigNFW}
\sigma^2(x=r/r_s)  = V_{\rm vir}^2 \frac{\cvir}{g(\cvir)} x(1+x)^2
\int_{x}^{\infty} \frac{g(x')}{x'^3(1+x')^2}{\mathrm{d}} x'.
\eeq
We find the following approximation useful and accurate to
$1\%$  for $x=0.01 - 100$:
\beq
\label{eq:sigrfit}
\sigma(x) \simeq V_{\rm max}\frac{1.4393x^{0.354}}{1 + 1.1756x^{0.725}}.
\eeq

There has been much debate on the appropriate way to assign the
Coulomb logarithm in Eq. (\ref{eq:Fdf}).  
Dynamical friction is caused by the scattering of
background particles into an overdense ``wake'' 
that trails the orbiting body and tugs back on 
the scatterer.  The
Coulomb logarithm  is interpreted as $\ln  (b_{\rm max}/b_{\rm min})$,
where $b_{\rm max}$ is the  maximum relevant impact parameter at which
background  particles  are  scattered into the wake  and
 $b_{\rm  min}$ is  the
minimum  relevant  impact parameter.   A  common  approximation is  to
choose  a  constant  value   of  the  Coulomb  logarithm  (perhaps  by
calibrating to the  results  of  numerical
experiments as in TB01), but some studies indicate that
this  approach significantly  underestimates
the dynamical friction timescale when tested against  N-body
simulations  (\eg Colpi, Mayer, \& Governato 1999; 
HFM02).  Motivated by the results of  HFM02 and 
Valenzuela \& Klypin (in preparation), we allow 
the  Coulomb logarithm to evolve with time 
and set $b_{\rm max}=r(t)$, where $r(t)$ is the radial  position  
of  the  orbiting  subhalo.  We assign the minimum 
impact parameter according to the prescription of White (1976) 
and integrate the  effect of encounters with background particles 
over the density profile of the subhalo.  Repeating this calculation for 
an NFW halo yields
\beq
\label{eq:lnL_exact}
\lnL = \ln \Bigg(\frac{r}{R_{\rm sat}} \Bigg) +
\frac{1}{g^2(x_{\rm sat})}I(x_{\rm sat}),
\eeq
where
\beq
\label{eq:Idef}
I(x_{\rm sat}) \equiv \int_{0}^{x_{\rm sat}} x_{\rm b}^3
\Bigg[\int_{x_{\rm b}}^{\infty} \frac{g(x)}{x^2 \sqrt{x^2 - x_{\rm b}^2}}
\textrm{d}x \Bigg]^2 \textrm{d}x_{\rm b},
\eeq
$x_{\rm sat} \equiv R_{\rm sat}/r_{\rm s}^{\rm sat}$, and 
$r_{\rm s}^{\rm sat}$ is the NFW scale radius of the satellite.   
The integral $I(y)$ is well-approximated by the following 
function, which is accurate to $1\%$ for $0.1 \le y \le 100$:
\beq
\label{eq:Ifit}
I(y) \simeq
\frac{0.10947 y^{3.989}}{ [ 1 + 0.90055y^{1.099} + 0.03568y^{1.189} + 0.06403y^{1.989} ] }.
\eeq

As the satellite orbits within the host potential, it 
is stripped of mass by the tidal forces that it experiences.  
First, we estimate the instantaneous tidal radius
of the subhalo $\rt$, at each point along its orbit.  
In the limit that the satellite is
much smaller than the host, the tidal radius is given by the 
solution to the equation (von Hoerner 1957; King 1962)
\beq
\label{eq:rtide}
\rt^3 \simeq \frac{\Msat(<\rt)/\Mhost(<r)}
{2 + \omega^2 R^3/G\Mhost(<r) - \partial \ln \Mhost(<r)/\partial \ln r}r^3,
\eeq
where $r$  is the radial position of the satellite, $\Mhost(<r)$  is the
host's  mass  contained  within  this  radius  [\cf Eq. (\ref{eq:MofR})],   
$M_{\rm   sat}(<r_{\rm  t})$   is   the
satellite's  mass contained within  $\rt$,  and $\omega$  is the
instantaneous angular speed of the satellite.  Equation (\ref{eq:rtide}) is 
merely an estimate of the satellite's tidal limit.  For a satellite on a 
circular orbit, it represents the distance from the satellite center to 
the point along the line connecting the satellite and the host halo center 
where the tidal force on a test particle just balances the attractive 
force of the satellite.  In reality, the tidal limit of a satellite cannot be represented 
by a spherical surface: some particles within $\rt$ will be unbound while 
others without $\rt$ may be bound.  Nevertheless, TB01 showed that this can serve 
as a very useful approximation.

As the tidal radius shrinks, unbound mass in the periphery 
is stripped.  Tidal forces are strongest, and
$\rt$ smallest, when the orbit reaches  pericenter; 
however, all of the mass outside of 
$\rt$ is not stripped instantaneously at each pericenter passage. 
Rather, mass is gradually lost from the satellite on a timescale
set by the orbital energy of the liberated particles.  
Johnston (1998), found that the typical energy scale of tidally stripped debris
is set by the change in the host halo potential on the length
scale of the orbiting satellite, $\epsilon \approx \rt \dd \Phi_{\rm host}(r)/\dd r$.
Particles on circular orbits of energy $E$ and
$E \pm \epsilon$ move a distance $\rt$ relative to each other on
a timescale of order the orbital period, $T$.  As such, we may
expect $T$ to be the relevant timescale for tidal mass stripping.  
TB01 used this timescale in their model to reproduce the results of 
several idealized N-body experiments.  Following TB01, we model 
satellite mass loss by dividing the orbit into discrete 
time steps of size $\delta t \ll T$.  At each step, we remove an amount 
of mass
\beq
\label{eq:dm}
\delta  m =  \Msat(>\rt)  \frac {\delta  t}{T}, 
\eeq  
where $\Msat(>\rt)$  is the  satellite's mass
exterior to $\rt$. 

As a subhalo loses mass due to tidal stripping, we assume that
its density profile is unmodified within its outer
radius $R_{\rm sat}$.  Rather than
identify $R_{\rm sat}$ with the tidal radius (which does not vary
monotonically with time), we set its value by determining the radius 
within which the mass profile retains the appropriate bound 
mass using Eq. (\ref{eq:MofR}).
We fix the scale radius of the subhalo $r_{\rm s}^{\rm sat}$, 
at the value defined at the epoch of accretion.   

Our approximations for dynamical friction  and tidal stripping are
least accurate  when the mass  of the  satellite is not very  small
compared  to the mass  of the host.  However, as $F_{\rm DF} \propto
\Msat^2$, it is in precisely these  cases that we expect the satellite
to merge quickly with the host and no longer  be identifiable  as
distinct substructure.  As such, the  precise dynamics should not have
a significant effect  upon our results  in these cases, particularly 
because our  main predictions involve low-mass substructure.  
However,  more detailed modeling will
be important for  investigations that focus on more massive substructures,
for example, explorations that use  disk thickening as  a  test of  the
$\Lambda$CDM cosmological model (\eg Font et al. 2001).

The final ingredients for our semi-analytic model of halo substructure
are the  criteria for  declaring subhalos to  be tidally  disrupted and 
centrally merged.   Let $\rmaxsat \simeq 2.16 \rssat$ be the radius at
which the subhalo's initial  velocity profile attains its maximum, and
$\Msat(<\rmaxsat)$ be  the mass of the  satellite originally contained
within the radius $\rmaxsat$.  We  declare a subhalo to be centrally
merged with the host if its  radial position relative to the center of
the  host becomes smaller  than $\rmaxsat$.   We  declare a
satellite tidally disrupted if the  mass of the satellite
becomes  less than  $\Msat(<\rmaxsat)$.  This  criterion  is partially
motivated by the numerical study of H03, who find
that NFW subhalos are completely tidally destroyed shortly after $\rt$
becomes less than $r_{\rm  max}^{\rm sat}$.  Of course the distinction
between  centrally merged  and  tidally destroyed  satellites  is
somewhat  arbitrary   as  subhalos  are   typically  severely  tidally
disrupted as they approach the  center of the host potential.
Fortunately, for the issues we explore here, the precise nature of
a satellite halo's destruction is not important.  We 
discuss  this  issue further  in  a  forthcoming  extension   of  our  work
(A. Zentner \& J. Bullock, in preparation).

In reporting results concerning the velocity function of substructure,
we  invoke  a  further modification.   H03  noted  that  subhalos  that
experienced significant tidal stripping suffered not only mass loss 
at radii $\gsim \rt$, but mass redistribution in their central regions, 
at radii smaller than  $\rt$.   To
account for this, we determine whether  or not the tidal radius of each
surviving subhalo was ever less than $\rmaxsat$.  If  so, we follow the 
prescription of H03 to account for mass redistribution and scale the
maximum circular velocity of the satellite via
\beq
\Vmax^{\rm final} = \Bigg(\frac{\Msat^{\rm final}}
{\Msat^{\rm initial}}\Bigg)^{1/3}\Vmax^{\rm initial},
\eeq
where $\Vmax^{\rm  initial}$ is the  maximum circular velocity  of the
satellite  according  to  its  initial  density  profile,  $\Msat^{\rm
final}$ is  its final mass,  and $\Msat^{\rm initial}$ is  its initial
mass  before being  tidally stripped.  In practice, this rescaling has 
a fairly small effect on our velocity functions.  Roughly $\sim 30\%$ of surviving 
halos meet this condition for $\rt$.  For those halos that do experience
this kind of mass loss, the typical reduction in $\Vmax$ is $\lsim 25\%$.

We are currently in the process of
checking  this model against   idealized   N-body
experiments designed to  mimic the type  of orbital histories that 
we encounter here (J. Bullock, K. Johnston, \& A. Zentner, in preparation).
Preliminary results show promising agreement.

\begin{figure}[t]
\begin{center}
\resizebox{!}{9.cm}
{\includegraphics{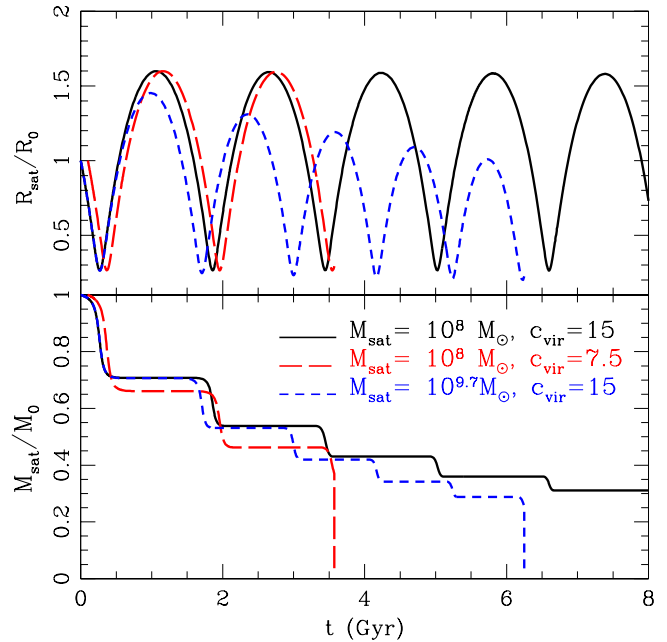}}
\figcaption{\label{fig:orbits} Orbital
    evolution  for three sets    of   subhalo parameters:
$M_{sat}^0=10^8$ \Msun,  $\cvir=15$ (solid); $M_{sat}^0=10^{8}$ \Msun, $\cvir=7.5$
(dashed); and $M_{sat}^0=5\times10^{9}$ \Msun, $\cvir=15$ (short-dashed).  Initial
orbital parameters and host mass properties are fixed, as described
in the text.  The top panel shows the radial evolution in units of the initial radius
as a function of time.  The bottom panel shows the mass of each system 
as a function of time.  Lines that terminate represent subhalo destruction at 
the end point (see text).}
\end{center}
\end{figure}

\begin{figure}[t]
\begin{center}
\resizebox{!}{9cm}
{\includegraphics{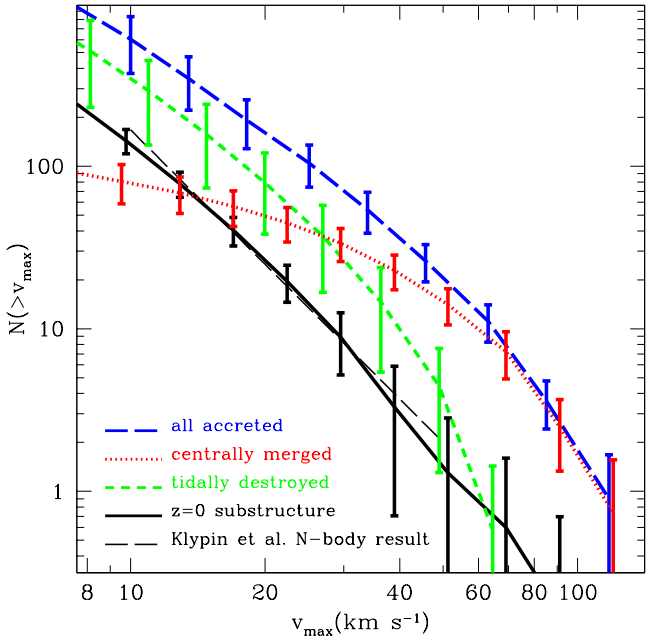}}
\caption{\label{fig:cal}The velocity functions of progenitor 
and surviving subhalo populations derived using our fiducial 
($n=1$, $\sigma_8=0.95$)
$\Lambda$CDM  cosmology and a 200-halo ensemble
of $1.4\times10^{12}$ \Msun systems at $z=0$.  Shown are all accreted
halos (dashed), and the fraction of those that are tidally
destroyed (short-dashed) and centrally merged (dotted).
The solid line shows the surviving population of subhalos
at $z=0$ and, for comparison, the thin dashed line shows 
the surviving population derived by K99 
using N-body simulations.  The error bars represent the sample
variance.}
\end{center}
\end{figure}
%
%
%
\subsection{\label{sub:test}Tests and Examples}
Figure \ref{fig:orbits} shows  three  example calculations of subhalo
trajectories  aimed  at demonstrating how   various factors affect the
orbital evolution of  a satellite system.   Each satellite was
started on the same initial orbit,   $\epsilon = \eta  = 0.5$, but
the satellite properties  were  varied:  $M_{\rm  sat}^0=10^8$ M$_{\odot}$,
$\cvir=15$  (solid); $M_{\rm sat}^0=10^{8}$ M$_{\odot}$, $\cvir=7.5$ (dashed); and
$M_{\rm sat}^0=5\times10^{9}$ M$_{\odot}$, $\cvir=15$ (short-dashed).  The  upper
and  lower panels depict the evolution  of orbital radius  and mass of
the subhalo respectively.  The  accretion time was  set at $8$ Gyr in
the past, $a=(1+z)^{-1} \simeq  0.45$ for this cosmology.  The host
halo parameters were chosen to match reasonable expectations for a
Milky Way-sized progenitor at that time: $M_{\rm
host}=5\times10^{11}$ \Msun ($R_{\rm vir}\simeq 110$kpc) and 
$c_{\rm  vir}=6$.   While the  subhalo represented  by  the solid line
experiences gradual  tidal  mass loss and  slight  orbital decay  as a
result of    dynamical  friction, its  core   survives  for  the  full
time period.  The less concentrated subhalo  (dashed) is more strongly
affected  by   tides, and is   completely   disrupted $\sim 3.5$  Gyr after
being incorporated into the host. (Although not shown, a similar effect is  seen if the host
halo  concentration is increased and the subhalo concentration is held
fixed.)  In the case of the massive subhalo, dynamical friction causes
the orbit to decay more quickly and the subhalo experiences more frequent
pericenter passages.  Consequently, disruption  occurs $\sim 6$ Gyr
after accretion.   Notice that because the  stripping process is gradual
(unless orbits are very radial) and the timescales involved are of 
order $\sim$ Gyr, the accretion time is also important
in determining survival probability.  If any of these subhalos were 
were accreted more recently, their chance of survival to  the present day would 
increase accordingly.  The  combination  of  factors  illustrated  here ---
accretion time, satellite mass, and the relative concentrations of
host and satellite  ---  will be important in later sections  for
understanding the factors that set the subhalo population from one 
cosmology to the next.
%
%
%

\begin{figure}[t]
\begin{center}
\resizebox{!}{9cm}
{\includegraphics{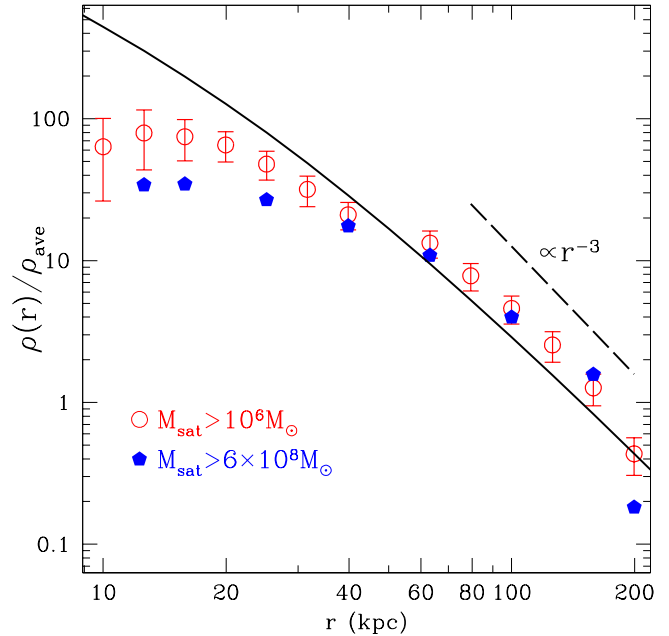}}
\caption{\label{fig:radial}The radial number density 
profile of substructure
derived from 200 model realizations of a $M_{\rm vir}=1.4\times10^{12}$
\Msun host halo at $z=0$ in our fiducial ($n=1$, $\sigma_8= 0.95$)
$\Lambda$CDM  cosmology.  The open circles show the  number density of subhalos
with $M_{\rm sat} > 10^{6}$ \Msun  divided by the average number density of
systems meeting this mass threshold within the virial radius of the
host system.  The points reflect the  radial profile averaged over all
realizations, and the error bars reflect the  sample variance.  
Solid  pentagons show the same result  for $M_{\rm sat} > 6\times
10^{8}$ \Msun  subhalos.  The variance (not shown) is significantly
larger for the higher mass threshold because there are significantly 
fewer such systems in each host.  For  reference, the solid line shows 
the NFW density profile of the host at $z=0$.   The virial radius for a
host halo of this size is $R_{\rm vir} \simeq 285$ kpc and the typical
NFW scale radius is $r_{\rm s} \simeq 20$ kpc.  We do not plot predictions 
beyond $r = 0.75 \Rvir \simeq 215$ kpc because this is the maximum 
circular radius we assign to in-falling, bound systems.}
\end{center}
\end{figure}
%
%
%

Figure \ref{fig:cal} shows the ensemble-averaged, cumulative velocity
function for the substructure population of Milky Way-like host halos
computed in our standard $\Lambda$CDM cosmology.  The host properties at
$z=0$ are $M_{\rm vir} =  1.4\times 10^{12}$ M$_{\odot}$, $c_{\rm vir}
\simeq 13.9$, and  $\Vmax \simeq 187 \kms$.   The lines  represent the
means of  200 merger tree  realizations, and the  error bars represent
the sample variances  over these  realizations.  
In particular, the thick solid line shows the surviving subhalo population
at $z=0$.   For  comparison, the   thin  dashed line is  the  best-fit
velocity  function reported  by K99 based  on an
analysis of substructure  in $\Lambda$CDM halos.  This line is  plotted over
the range that their resolution and sample size allowed them to probe.
The apparent agreement between our  semi-analytic model and the N-body
result is excellent, and lends confidence in our ability to apply this
model to different power spectra.

The radial distribution of substructure at $z=0$ for the same
ensemble of halos is shown in Figure \ref{fig:radial}.  Open circles
show the differential number density profile of subhalos with $M_{\rm
sat}>10^{6}$ \Msun normalized relative to the total, volume-averaged
number density of subhalos within $\Rvir$ that meet the same mass requirement.
The solid pentagons  show  the same quantity for
more massive subhalos, $M_{\rm sat}>6\times 10^{8}$ M$_{\odot}$. The line
shows the NFW dark matter profile for the host system normalized
relative to the average (virial) density  within the halo.  Observe that
the  subhalo profile traces the  dark matter profile  at large radius 
($\rho \propto r^{-3}$), but flattens  towards  the center as a consequence 
of tidal disruption.  This result  agrees well 
with that presented  in Figure  3 of  Col{\'{\i}}n et
al. (1999).  Using an N-body analysis of a  cluster-sized host,
Col{\'{\i}}n et al. (1999) showed that the number density of
systems with $M_{\rm sat}$ greater than $0.04\%$ of the host mass 
traces the  background  halo profile at  large  radius,
begins to flatten at $r \sim 0.2 \Rvir$, and is 
roughly a factor of $5$ below the background at $r \sim 0.07 \Rvir$
(their innermost point).\footnote{We quote results relative to $\Rvir$ and 
$M_{\rm host}$ because the host halo in Col{\'{\i}}n et al. (1999) is 
significantly more massive than the halos that we consider.}  
The solid pentagons in Figure  \ref{fig:radial} correspond to the same
mass fraction relative  to the host.   Notice that at  $r = 0.07 \Rvir
\simeq 20$ kpc, the factor of $\sim 5$ mismatch is reproduced.  Ghigna
et al. (1998) observed the same qualitative behavior for subhalos in a
standard CDM simulation  of a cluster-size  halo.  Chen et  al. (2003)
have   measured  the   substructure  profile  using a  high-resolution
galaxy-size  halo with $M_{\rm  sat}   \gsim 0.0015\% M_{\rm  host}$,
corresponding   to subhalos   intermediate   in  mass   between  those
represented  by the open circles  and  solid pentagons  in Figure
\ref{fig:radial}.  Chen et  al.  (2003) similarly find  core  behavior
setting in at  a radius of  $\sim 30$  kpc,  but also find a  stronger
overall  suppression in substructure  counts within $r  \lsim 70$ kpc.
Our  results  suggest that  some of the   observed suppression  may be
caused  by overmerging in the central regions of their simulated halos.  
Ongoing studies by other workers lead to similar conclusions 
(J. Taylor, private communication).  
Only the next generation of numerical simulations can reliably
test this.  That we produce a reasonable approximation to the 
number density profile of substructure is an indication of the 
soundness of our model.  

\section{\label{sec:ps}Model Power Spectra}

The  initial   power  spectrum of  density fluctuations is conventionally 
written  as an approximate power  law  in wavenumber $k$, $P(k)  \propto
k^n$, corresponding to a variance per logarithmic interval in
wavenumber of $\Delta^2(k) = k^3P(k)/2\pi^2$.  If the 
fluctuations were seeded during an early inflationary stage, as is  
commonly supposed, then  the initial spectrum  is  likely  to  be nearly 
scale-invariant,  with  $n \simeq  1$.  Any deviation from power law behavior, 
or ``running''  of the power law index with scale is likewise expected to be small, 
$|\dd n / \dd \ln k| < 0.01$ (Kosowsky \& Turner 1995).  In
addition   to these  theoretical prejudices, large-scale observations 
of galaxy clustering and CMB anisotropy seem to favor nearly 
scale-invariant models that can be parameterized in this way.  In this 
paper we explore the effects on halo substructure of taking $n \ne 1$ and 
allowing for scale-dependence in the power law index and 
more dramatic features in the power spectrum.  In this 
section we give a brief description of the power spectra that we explore.  

Table \ref{table:spectra} summarizes the relevant features of our example 
power spectra.  The second and third  columns
list the primordial spectral index evaluated
at the pivot scale of the COBE measurements $k_{\rm  COBE} \approx 0.0023$ 
h Mpc$^{-1}$, and the running of the spectral index.\footnote{We use the 
definition of running employed by Spergel et al. (2003) rather than that 
given in, for instance, Kosowsky \& Turner (1995).  These definitions differ by a factor of two.}  
We neglect any variation in the running with scale and explicitly 
set $\textrm{d}^2 n(k)/\textrm{d} (\ln k)^2 = 0$.  
Except for the running index (RI) case, we normalize all models 
to the COBE measurements of the CMB anisotropy using the fitting formulae 
of Bunn, Liddle, \& White (1996; also  Bunn \& White 1997).  The fourth 
column of Table \ref{table:spectra} gives the implied value of $\sigma_8$.  
We calculate spectra using the transfer functions of Eisenstein \&  Hu (1999).
In Figure \ref{fig:spectra} we illustrate the implied $\sigma(M)$ 
for these models.

Many of the spectra listed in Table \ref{table:spectra} 
are motivated by particular models of inflation.  We invoke an inverse power 
law potential that gives rise to a mild tilt $n\simeq 0.94$, as well as 
a model in which the inflaton has a logarithmically running mass and which 
can give rise to significant tilt and running for natural 
parameter choices (Stewart 1997a,b; Covi \& Lyth 1999; Covi, Lyth, 
\& Roszkowski 1999; Covi, Lyth, \& Melchiorri 2003).  We employ specific 
inflationary potentials   mainly as a  conceptual follow  up  to ZB02,
which highlighted the   fact  that various  levels of   tilt may occur
naturally within the paradigm of inflation and that $n\equiv 1$ is not
demanded  by this paradigm.   For the purposes of  this paper, one may
regard our choices simply as spanning a range of observationally viable 
input  power spectra.  The values of tilt and $\sig8$ that we consider range 
from  $n \simeq  0.84$ with $\sig8=0.65$ to $n =  1$ and $\sig8 = 0.95$.  
The model with $\sig8 = 0.75$ was  specifically chosen to  match galaxy 
central densities, as described in ZB02.  We also explore the best-fit, 
running-index model of the WMAP team (Spergel et al. 2003), 
with ${\rm d}n/{\rm d}\ln k = -0.03$. We refer to this as the 
``running index model'' or ``RI model.''  Note that Spergel et al. (2003) 
quote a value of $n =0.93$ evaluated at $k=0.05$ Mpc$^{-1}$.  The  value listed in Table 1
is larger because we quote it at a smaller wavenumber, $k=k_{\rm COBE}$. 

In addition to tilted $\Lambda$CDM models, we 
consider spectra with abrupt reductions in power on small
scales.  In the ``broken scale-invariance'' (BSI) example, we adopt an
idealized inflation model introduced by Starobinsky  (1992) that exhibits 
the most rapid drop in power possible for a single field model.
Kamionkowski \& Liddle (2000) studied this type of model as a way 
to mitigate the dwarf  satellite problem, but our
choice of parameters is slightly different from theirs (see ZB02).

We also consider WDM scenarios in which the primordial power spectra are 
scale-invariant but small-scale fluctuations are subsequently filtered by 
free-streaming.  The free-streaming scale is set by the 
primordial velocity dispersion of the warm particles.  In the 
canonical case of a ``neutrino-like,'' thermal relic with two internal 
degrees of freedom, the free-streaming scale 
can be expressed in terms of the warm particle mass $m_{\rm W}$ 
and relic abundance, $\Omega_{\rm W}h^2$:

\beq
\label{eq:Rf}
R_f \simeq 0.11 \left[\frac{\Ow h^2}{0.15} \right]^{1/3}
\left[ \frac{m_{\rm W}}{{\rm keV}} \right]^{-4/3}{\rm Mpc}.
\eeq
We calculate WDM spectra assuming the same flat cosmology with
 $\Omega_{\rm M}=\Ow + \Omega_{\rm B} = 0.3$, and use the approximate 
WDM transfer function given by Bardeen et al. (1986), 
$P(k) = \exp[-k R_f - (k R_f)^2] P_{\rm CDM}(k)$.

Several studies have placed approximate constraints on WDM masses based
on either the argument that there must be enough power on small scales
to reionize the Universe at sufficiently high redshift 
($z_{\rm re} \gsim 6$) or  by probing the  power spectrum on  small scales directly
with the Ly-$\alpha$ forest (Barkana, Haiman, \&  Ostriker  2001;
Narayan  et  al. 2000).  These  authors essentially  find that 
$m_{\rm {\rm W}} \gsim 0.75$ keV assuming a  neutrino-like thermal relic; 
however, this constraint may be significantly
more restrictive if measurements of $z_{\rm  re} \sim 17$ by the WMAP 
collaboration (Kogut  et al.  2003; Spergel et al. 2003) are confirmed 
(Somerville, Bullock, \& Livio 2003).  As  such, we consider three
illustrative examples in what follows, $m_{\rm W} = 0.75$ keV, $1.5$ keV, and 
$3.0$ keV.  The corresponding ``free-streaming'' masses, below which the fluctuation
amplitudes are suppressed, are listed in Table \ref{table:spectra}.

%
%
\section{\label{sec:results}RESULTS}
%
%
\subsection{\label{sub:mh}Accretion Histories}

Our first results concern the merger histories of halos that are 
approximately Milky Way-sized, with $M_{\rm vir} = 1.4 \times   10^{12}$ 
\Msun at  $z=0$.  For the $n=1$, $\Lambda$CDM model, we present results 
based on 200 realizations.  For all other models, our findings are based 
on 50 model realizations.  

%
%
%
%
\onecolumn
%
\begin{deluxetable}{lcccc}
\tablecaption{\label{table:spectra}Summary of power spectra properties}
\tablehead{
\colhead{Model Description} &  \colhead{$n(k_{\rm COBE})$} 
& \colhead{$\textrm{d}n(k)/\textrm{d}\ln k$} & \colhead{$\sigma_{8}$} & \colhead{comments} 
}
\startdata
Scale-invariant & $1.00$ & \hspace{.2cm}$0.000$ & $0.95$ \\
\\
Inverted power law inflation & $0.94$ & $ -0.002 $ & $0.83$ \\
Running-mass inflation I &  $ 0.84$ & $-0.008 $ & $0.65$ & see Stewart 1997a,b\\
Running-mass inflation II & $0.90 $ & $-0.002$ & $0.75$ \\
WMAP best-fit running index (RI) model & $1.03$   & $-0.03$   & $0.84$ & WMAP best fit, see Spergel et al. 2003\\ 
Broken scale-invariant inflation (BSI) & $1.00$ &  \hspace{.2cm}$ 0.000$ & $0.97$ & 
exhibits sharp decline in power at $k \gsim 1$ h Mpc$^{-1}$,\\ 
 & & & & power suppressed for $M \lsim 10^{10}$ \Msun\\
\\
Warm Dark Matter, $m_{\rm W}=3.0$ keV & $1.00$ & \hspace{.2cm}$ 0.000$ & $0.95$ & $M_f  \simeq 8.3 \times 10^8$ \Msun\\
Warm Dark Matter, $m_{\rm W}=1.5$ keV& $1.00$ & \hspace{.2cm}$ 0.000$ & $0.95$ & $M_f  \simeq 1.3 \times 10^{10}$ \Msun\\
Warm Dark Matter, $m_{\rm W}=0.75$ keV& $1.00$ & \hspace{.2cm}$ 0.000$ & $0.94$ & $M_f  \simeq 2.1 \times 10^{11}$ \Msun\\
\enddata
\tablecomments{Column (1) gives a brief description  of the inflation or warm dark matter 
model used to predict the power spectrum.  In the text, we distinguish the first 
five models by their tilts and/or their values of $\sigma_8$.  We label 
the warm dark matter models by the warm particle mass.  Columns (2)  and (3)
give the tilt $n(k_{\rm COBE})$ on the pivot scale of the COBE data 
$k_{\rm COBE} \approx 0.0023$ h Mpc$^{-1}$, and the running of the spectral 
index $\textrm{d}n(k)/\textrm{d}\ln k$, respectively.  We have explicitly assumed the 
``running-of-running'' to be small and taken 
$\textrm{d}^2 n(k)/\textrm{d} (\ln k)^2 = 0$.  
Column (4) contains the values  of $\sigma_8$ implied by the tilt or warm particle 
mass, the COBE normalization, and our fiducial cosmological parameters except in 
the case of the WMAP best-fit running index (RI) model, in which case the value of 
$\sigma_8$ reflects their best-fit normalization.}
\end{deluxetable}
%
%
%
%
%
%
%
%
\begin{figure*}[t]
\centerline{ 
   \epsfysize=3.5truein  \epsffile{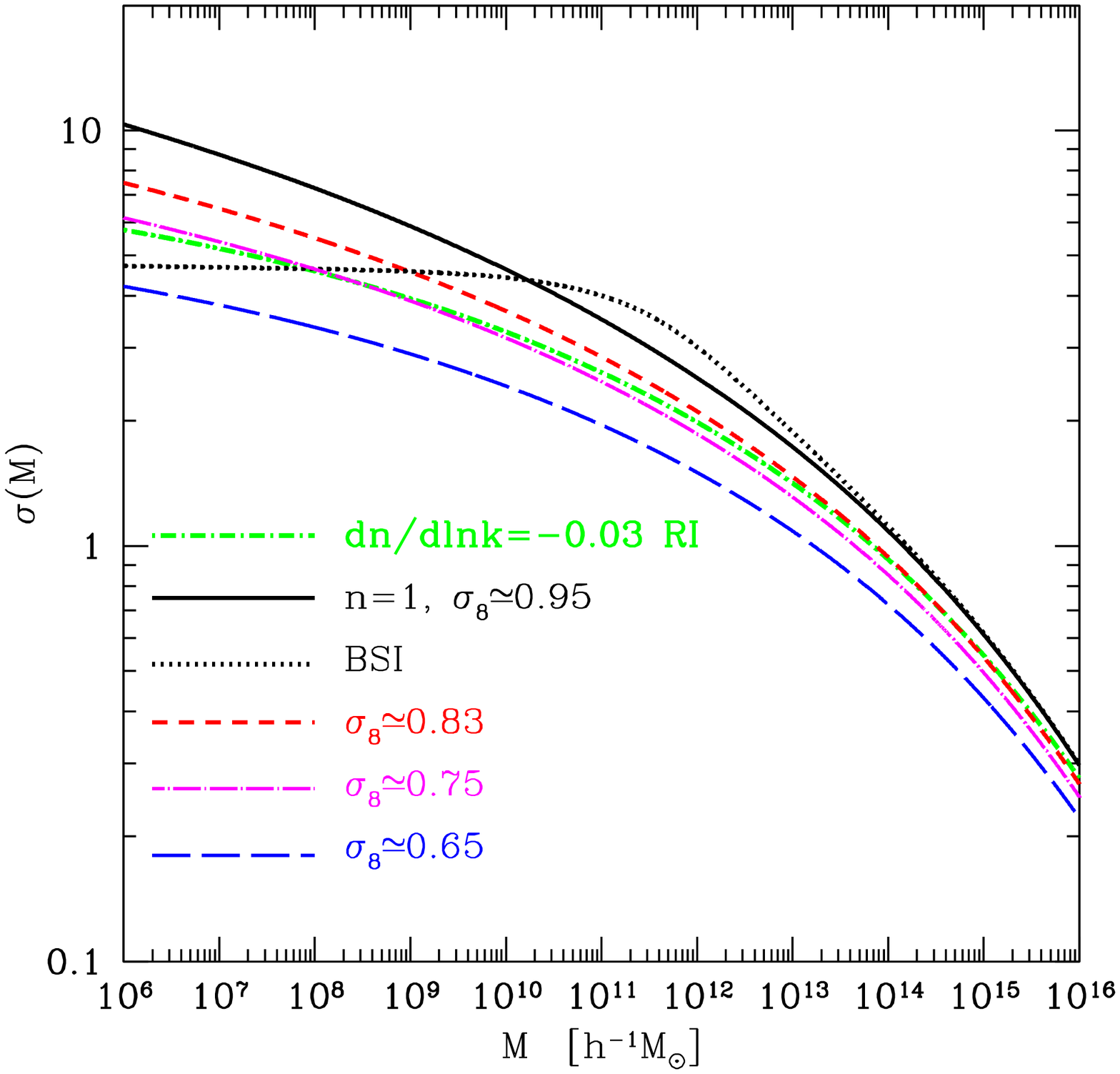}
   \epsfysize=3.5truein  \epsffile{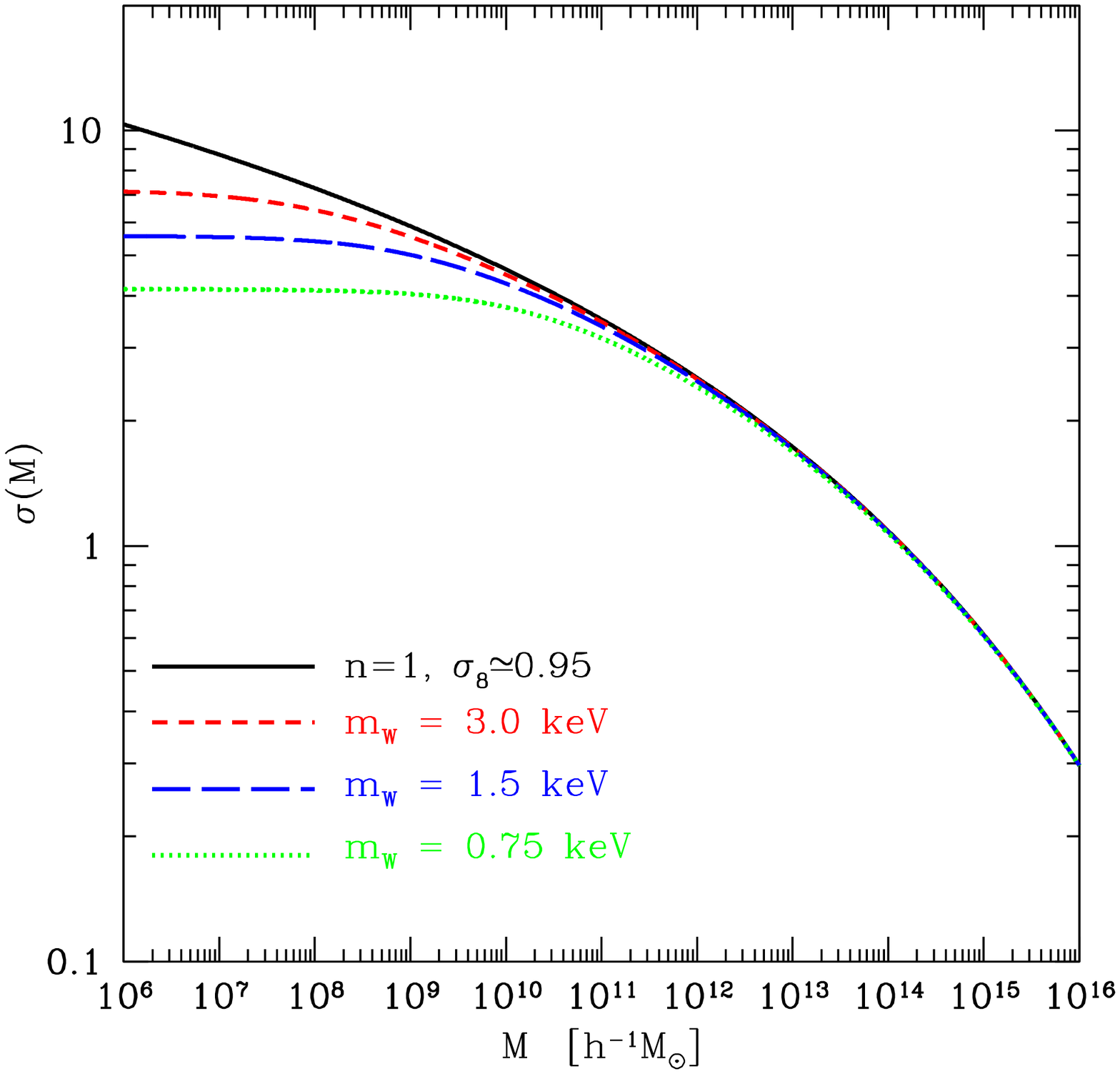}
}                                                  
\figcaption{\label{fig:spectra}The $z=0$ rms overdensity as a function of mass scale
for several of our adopted power spectra (see \S \ref{sec:ps} and Table \ref{table:spectra} 
for more details).  In the left panel, we exhibit spectra that deviate from the standard 
$n=1$, scale-invariant model.  The models shown in this panel are standard $n=1$ (solid), a broken
scale-invariant model (dotted),  $\sigma_8 = 0.84$ \& $n \simeq 0.94$ (short-dashed), 
$\sigma_8 = 0.65$ \& $n \simeq 0.84$ (long-dashed),  $\sigma_8 = 0.75$ \& $n \simeq 0.90$
(dot-long-dashed), and a model based on the results of the WMAP team with $n \simeq 1.03$, 
$\dd n / \dd \ln k = -0.03$, and $\sigma_8 = 0.84$ (dot-short-dashed).  
The inflation models that inspire these 
examples are described in ZB02.  In the right panel, we show several warm dark matter power 
spectra.  More precisely, we depict spectra implied by warm particle masses of 
$m_{\rm W} = 3.0$ keV (short-dashed), $m_{\rm W} = 1.5$ keV (long-dashed), and 
$m_{\rm W} = 0.75$ keV (dotted) along side the standard $n=1$, $\Lambda$CDM 
spectrum (solid).}
\end{figure*}
\twocolumn
%
%

\begin{figure}[t]
\begin{center}
\resizebox{!}{9cm}
{\includegraphics{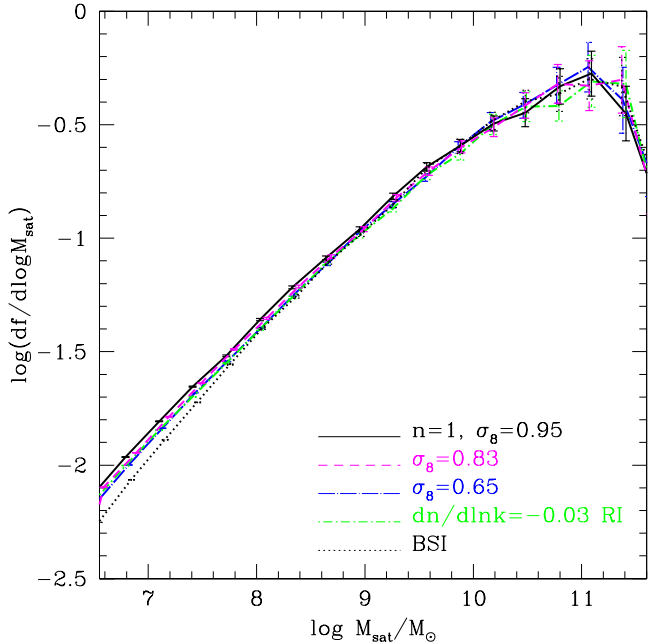}}
\caption{\label{fig:mergefrac} Fraction of final host mass 
accreted in subhalos of mass $M_{\rm sat}$ as a function of $M_{\rm sat}$.  
The final host mass is $1.4\times10^{12}$ M$_{\odot}$.  The results
for several input power spectra are shown.}
\end{center}
\end{figure}
%
%
%

Figure \ref{fig:mergefrac} focuses on  the  the mass distribution  of
accreted halos, integrated over the entire merger history of the host.
We plot $\dd f/ \dd \log (M_{\rm sat})$, the fraction of
mass in the final halo that was accreted in subhalos of a given mass 
per  logarithmic interval in subhalo  mass, $M_{\rm sat}$.  Observe that 
the mass fraction accreted in subhalos of a given 
mass is relatively insensitive to the shape of the power spectrum.
Although similarity from model to model may be somewhat surprising at first, 
it follows directly  from repeated application of Equation
(\ref{eq:merger}).  In  particular, the shape of  the progenitor 
distribution for $M_{\rm sat} \ll  M_{\rm host}$ must 
follow $\dd f/\dd \log(M_{\rm sat}) \propto M_{\rm sat}^{1/2}$, and
the turnover occurs because  mass conservation suppresses the
number of major mergers.  The shape shown in Figure
\ref{fig:mergefrac} and its insensitivity to the power 
spectrum is  discussed in detail by LC93.

\begin{figure}[t]
\begin{center}
\resizebox{!}{9cm}
{\includegraphics{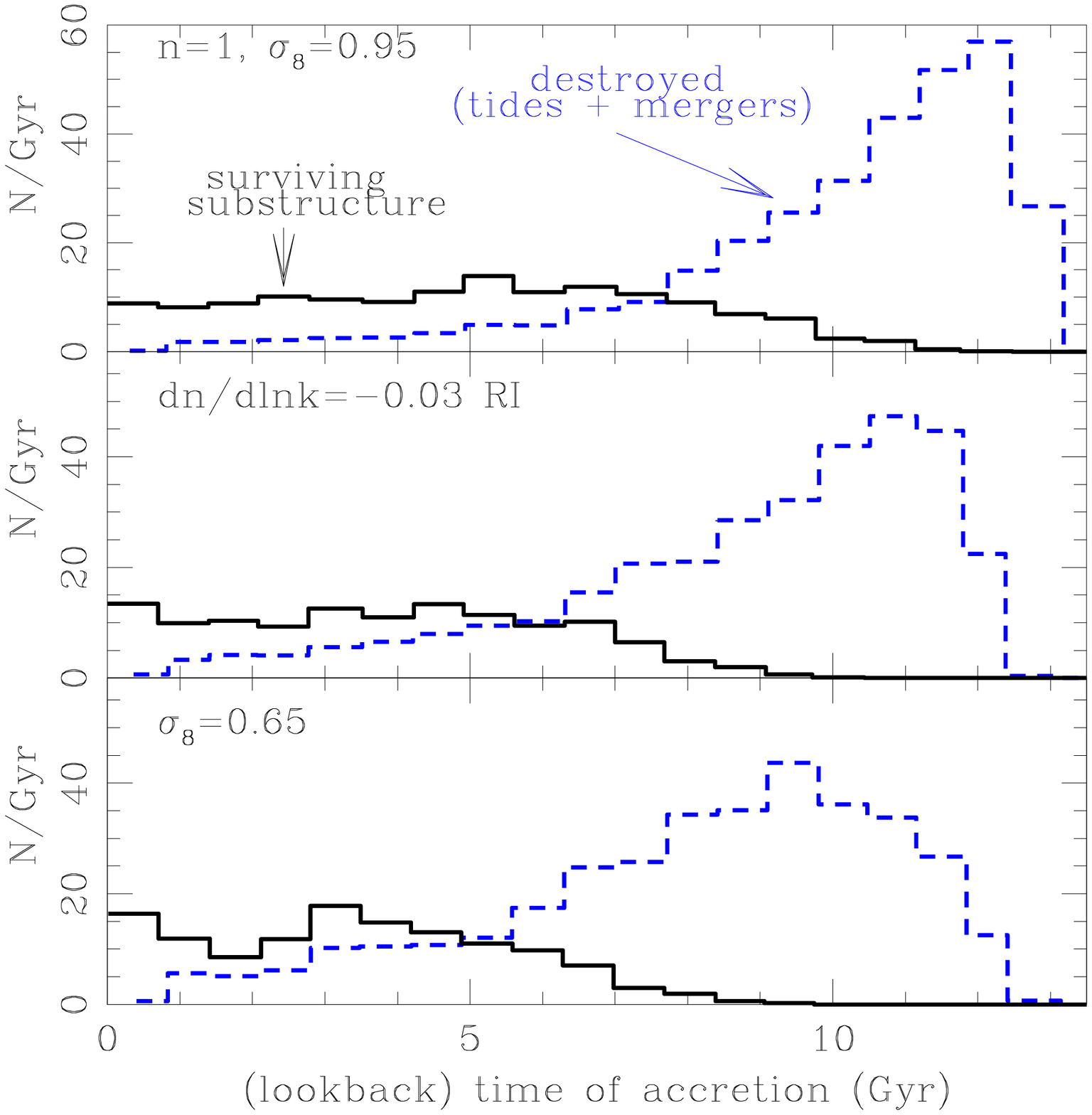}}
\caption{\label{fig:rate}
Accretion rate averages, $dN/dt$ (Gyr$^{-1}$), for merged halos
more massive than $10^8$ \Msun and host halos of
mass $1.4\times10^{12}$ \Msun at $z=0$.  Three different
power spectra are shown: $n=1$ (upper); RI model (middle); and $n=0.84$, 
$\sig8=0.65$ (bottom).  Dashed lines show objects that are destined to
be destroyed, either by tidal disruption or central merging, and
solid lines show subhalos that survive until $z=0$.  The $n=1$ results 
are based on 200 EPS realizations while the others are based on 50 
realizations.}
\end{center}
\end{figure}

While the  total mass function of  accreted substructure is relatively
independent of the power spectrum,  the merger histories
themselves are not.  In models with less power on galaxy scales, halos  
assemble their mass later and experience more recent mergers and
disruption events.
We show an example of this in Figure \ref{fig:rate}.  Here we plot the 
average accretion  rate of subhalos  with $M_{\rm sat} > 10^8$ \Msun 
for host halos in the standard $n=1$, $\Lambda$CDM  model, the RI model,
and our lowest normalization case ($n=0.84$, $\sig8 = 0.65$).  
The total accretion rate is divided in two pieces: dashed lines show those 
subhalos that are eventually destroyed and solid lines  show the accretion 
times of subhalos that survive until $z=0$.  
For the standard ($n=1$, $\sigma_8 =  0.95$) case, the event rate peaks 
sharply about $\sim 12$ Gyr in the past, while the $\sigma_8=0.65$ case 
has a broader distribution, peaking later at $\sim 9$ Gyr ago, and with a 
long tail of accretion events extending towards the present day.
%
%
%
%

%

%
%
%
\begin{figure}[t]
\begin{center}
\resizebox{!}{9cm}
{\includegraphics{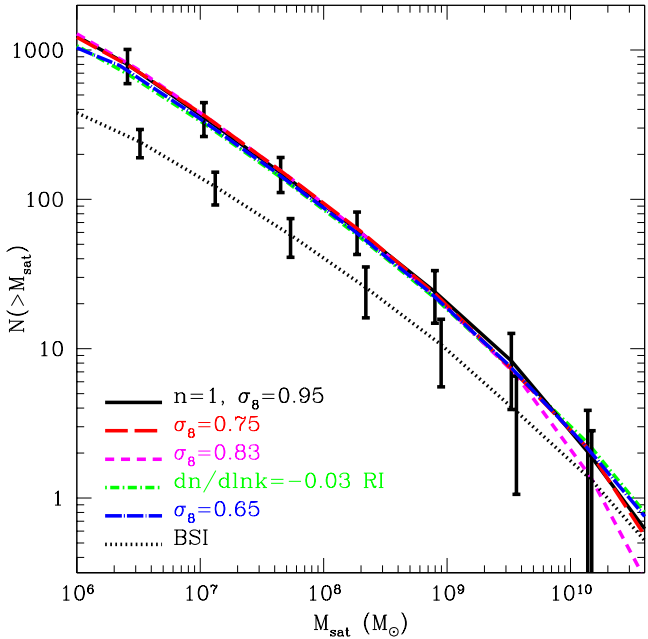}}
\caption{\label{fig:mfunc}The cumulative 
mass function $N(>M)$, of surviving subhalos computed for
an ensemble of host halos of mass 
$M_{\rm host} = 1.4 \times 10^{12} M_{\odot}$ at $z=0$.
The lines show the means computed over all realizations.  The 
different line types relate to different models following the
convention in the left panel of Figure \ref{fig:spectra}.  
The $n=1$ results are derived from 200 realizations and the 
results for the other models are based on 50 realizations.  
The error bars show the variance over the $n=1$ realizations (upper)
and BSI realizations (lower).}
\end{center}
\end{figure} 
%
%
%
%

\begin{figure}[t!]
\begin{center}
\resizebox{!}{9cm}
{\includegraphics{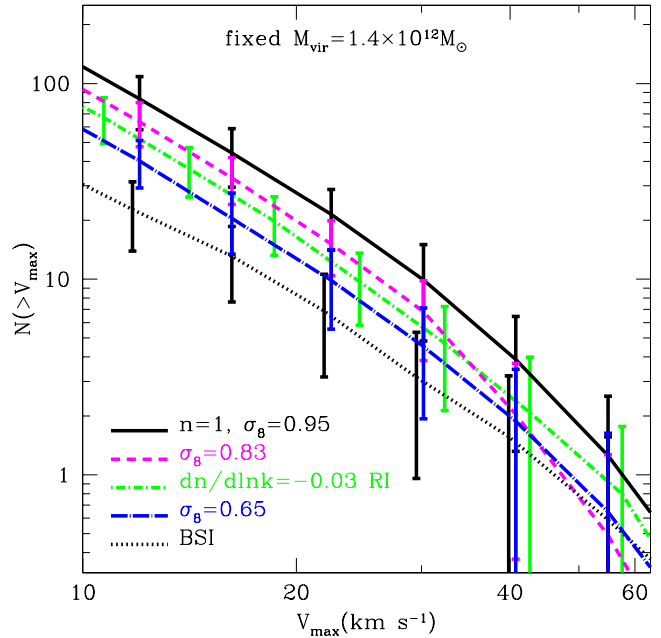}}
\caption{\label{fig:vcomp}The cumulative velocity function 
of subhalos in a host of fixed mass, $\Mhost = 1.4 \times 10^{12}$ 
M$_{\odot}$.  The lines represent averages over 200 merger 
histories for $n=1$ and 50 merger histories for the other models.  
The error bars represent the dispersion amongst 
these realizations.  The models are $n=1$ (solid), $\sig8 \simeq 0.83$ (dashed), 
RI model (dot-short-dashed), $\sig8 \simeq 0.65$ (dot-long-dashed), and 
BSI (dotted).}
\end{center}
\end{figure}
%
%

The shift in accretion times in models with less small-scale power plays an 
important role in regulating the number of surviving subhalos.  As we 
discussed in relation  to Figure \ref{fig:orbits}, 
a finite  amount of time is required for an orbit to  decay or for 
a system  to become unbound and in many cases 
the longer a subhalo orbits in the background potential, the more
probable its disruption becomes.  The later  accretion times in models
with less power partially  compensate for the fact that subhalos
in these  models  are less centrally  concentrated and  more
susceptible to disruption at each pericenter passage.  Particular 
results for substructure populations in each  model are given in
the following subsections.

That we expect a characteristic merger/disruption phase in each halo's
past is  intriguing, as this phase is approximately coincident with  the
estimated ages of galactic thick disks, $t_{td} \sim  8-10$ Gyr
(\eg Quillen \& Garnett 2000 for the Milky Way), which seem to be
ubiquitous and roughly coeval (Dalcanton \& Bernstein 2002).
In this context, the age distributions  of thick disks  might serve as a 
test of this characteristic accretion time, which varies as a
function  of normalization {\em and} cosmology.  We reiterate that
the look-back times shown for the dashed lines in Figure \ref{fig:rate}
are the times that the subhalos were {\em accreted}.  The distributions of 
central merger rates and tidal destruction rates peak at slightly more 
recent times and their widths are broader, 
with longer tails towards the present epoch.

It is interesting to note that the  surviving  halos in Figure \ref{fig:rate}
represent a distinctly different  population of objects  than  the
destroyed systems ---  they tend to have  been accreted more recently.
We are inclined to speculate that the  star formation  histories of
galaxies that were destroyed  after being accreted could be distinctly 
different from those of the surviving (dwarf satellite) galaxies as well.  
This may have implications for understanding whether
the  stellar halo of  our Galaxy formed  from disrupted dwarfs or some
other  process.  While the global structure  of the stellar halo seems
consistent with the  disruption  theory (Bullock,  Kravtsov,  \&
Weinberg 2001), the element ratios of stellar halo  stars and stars in
dwarf galaxies are not consistent with a common history of chemical 
evolution (Shetrone, C{\^o}te, \& Sargent 2001).  The results shown in Figure
\ref{fig:rate} provide general motivation to model dwarf galaxy evolution and 
Milky Way formation in a cosmological context.

%
%
%
\subsection{\label{sub:vfunc}Mass and Velocity Functions}

We present our results on CDM halo substructure beginning with the
abundance of satellites in Milky Way-like galaxies.  We plot the mass
function of subhalos $N(>M_{\rm sat})$, or  the  number of subhalos with mass
greater than $M_{\rm sat}$ as a function of $M_{\rm sat}$, for each of our  models in
Figure \ref{fig:mfunc}.   The host halo  mass  is again fixed  at $1.4
\times 10^{12}$ \Msun at $z=0$.  From  this figure, we  see that even in
the significantly  tilted,  low-normalization model  ($\sigma_8 \simeq
0.65$),  the number of satellite   halos with mass  greater than $10^6$ M$_{\odot}$ 
is roughly equal to that in the standard
$n=1$ model.  The systematic differences between models are 
small compared to the scatter.  The suppression is weak because 
several competing effects tend to compensate for the reduced 
concentrations of the subhalos.  In tilted models with reduced small-scale 
power, subhalos are typically accreted at later times.  In addition, host 
halos are less concentrated and correspondlingly less capable of 
disrupting their satellites.

In contrast, the BSI model shows a  substantial decrease (a
factor  of $\sim  3$) in the   number of surviving  satellite halos at
fixed  mass.  The reason  for the dramatic reduction   in this case is
easy to understand.  First, power is  reduced  only on scales  smaller
than a critical scale around  $\sim  10^{10}$ M$_{\odot}$ (\cf Figure
\ref{fig:spectra}) and so, the concentration and accretion history of
the $\sim 10^{12}$ M$_{\odot}$  host halo are minimally altered  while
the concentrations of the  small subhalos are drastically reduced (see
ZB02).  In other words, the host halo has a density structure similar 
the $n=1$ model host and is just as capable of tidally disrupting
satellites, but the satellites are significantly more susceptible to
disruption.   A second difference is that galaxy-size halos in the BSI
model, unlike the tilted models,  accrete  $\sim 40  \%$ fewer
low-mass ($\lsim 10^7$  \Msun) halos over their lifetimes, and this
further widens the disparity between the BSI and tilted-$\Lambda$CDM 
models.

It is conventional to discuss the substructure population of Milky
Way-like halos in terms of the velocity function.  In Figure
\ref{fig:vcomp}, we  show   our results for  the   cumulative velocity
functions of subhalos for a fixed host  mass of  $\Mhost = 1.4
\times 10^{12}$ M$_{\odot}$.  Notice  that the velocity functions show a
stronger trend with power spectrum than the mass functions (Figure
\ref{fig:mfunc}), but the effect is still modest compared to the 
statistical scatter.  For the most extreme
tilted model, the total number of subhalos with  $\Vmax \gsim 10 \kms$
is only   a    factor of   $\sim  2$  lower   than   in  the  standard, 
scale-invariant case.  In the case of the tilted models, the reduction
in the velocity function is largely due to the fact that the subhalos are 
less concentrated, so the $\Vmax$ values are correspondingly smaller for fixed 
halo masses [\cf Eq. (\ref{eq:MofR}) and the discussion that follows].

\begin{figure}[t]
\begin{center}
\resizebox{!}{9cm}
{\includegraphics{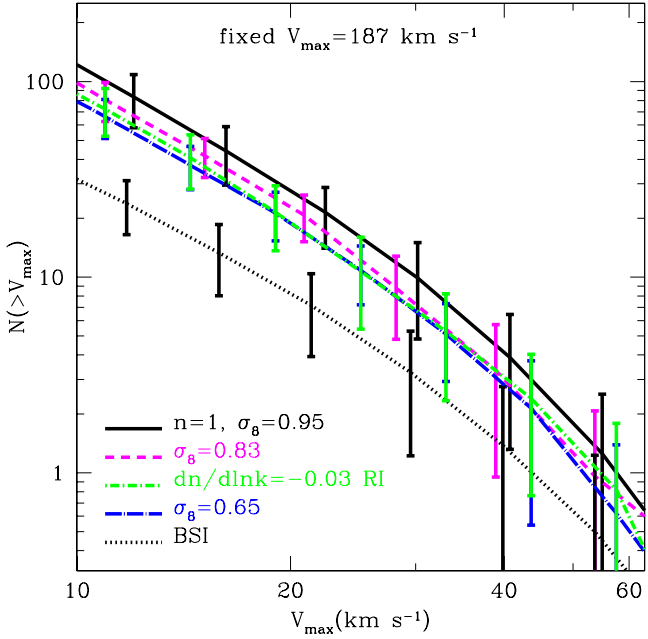}}
\caption{\label{fig:fixedvmax}The cumulative velocity function 
of subhalos in a host of fixed maximum velocity $\Vmax \simeq 187 \kms$.  
The lines represent averages over 200 merger history 
realizations for $n=1$ and 50 realizations for all other models.  
The error bars represent the dispersion in these realizations.  
The different models are as in Figure \ref{fig:vcomp}. }
\end{center}
\end{figure}
%
%
%
%
%
\begin{figure}[t]
\begin{center}
\resizebox{!}{9cm}
{\includegraphics{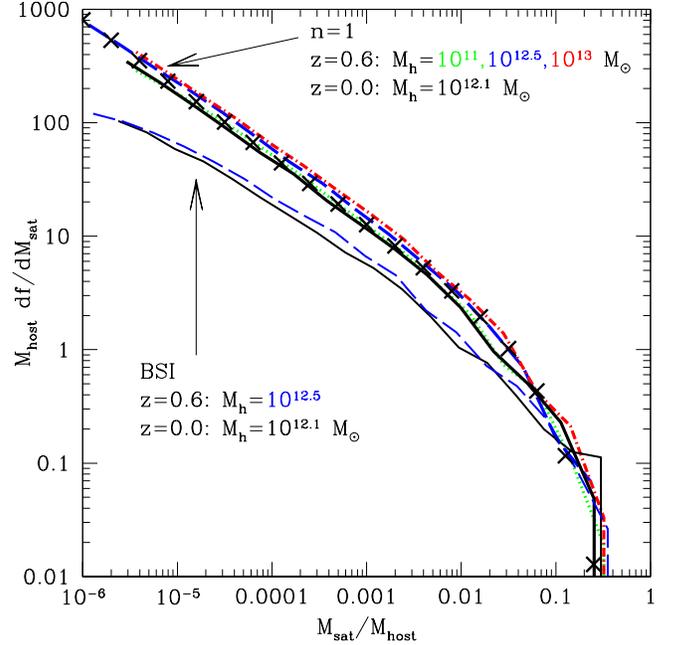}}
\caption{\label{fig:dfdm}The average differential
mass fraction, $df/dM_{\rm sat}$, 
normalized relative to host mass and satellite mass.
The upper set of (bold) curves were computed for the $n=1$ cosmology with
$M_{\rm host} = 1.4 \times 10^{12}$ M$_{\odot}$ at $z=0$ (solid)
and $M_{\rm host}=$ $10^{11}$ \Msun (dotted), $3\times10^{12}$
\Msun (long-dash) and $10^{13}$ \Msun (dot-dash) all at $z=0.6$.
The lower set of thin curves correspond to BSI halos of
$M_{\rm host} = 1.4 \times 10^{12}$ M$_{\odot}$ at $z=0$ (solid)
and $M_{\rm host}=3\times10^{12}$ \Msun at $z=0.6$ (long-dash).
The crosses reflect an analytic fit to the $n=1$ results, as discussed
in the text [see Eq. (\ref{eq:fitfrac})].}
\end{center}
\end{figure} 
%
%
%
%
%
\begin{figure}[t]
\begin{center}
\resizebox{!}{9cm}
{\includegraphics{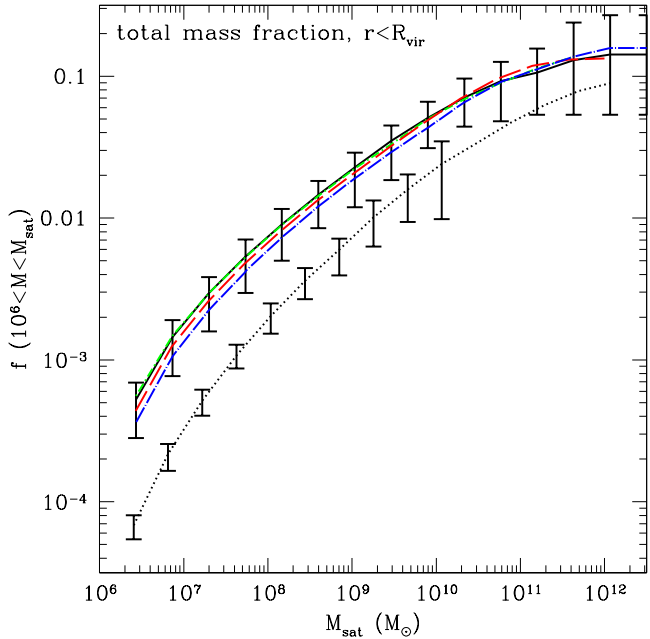}}
\caption{\label{fig:mfrac_total}The fraction of the 
parent halo mass that is bound up in substructure 
in the mass range between $10^6$ M$_{\odot}$ and 
$M_{\rm sat}$ as a function of $M_{\rm sat}$. The host halo in each case 
has $M=3\times10^{12}$ \Msun at $z=0.6$.
Lines reflect the mean over all realizations, and
results are shown for
the $n=1$ model (solid), RI model (dot-short-dash), 
 $\sig8=0.75$ (dashed), $\sig8=0.65$ (dot-long-dash), and
BSI (dotted).  The error bars on the top set of lines reflect
the $90$ percentile range determined using $200$ merger tree
realizations for the $n=1$ case (the other models in the top
set of lines have very similar scatter).  The bottom set of
errors reflect the same range determined using $50$ realizations
of the BSI model.}
\end{center}
\end{figure}
%
%
%
\begin{figure}[t]
\begin{center}
\resizebox{!}{9cm}
{\includegraphics{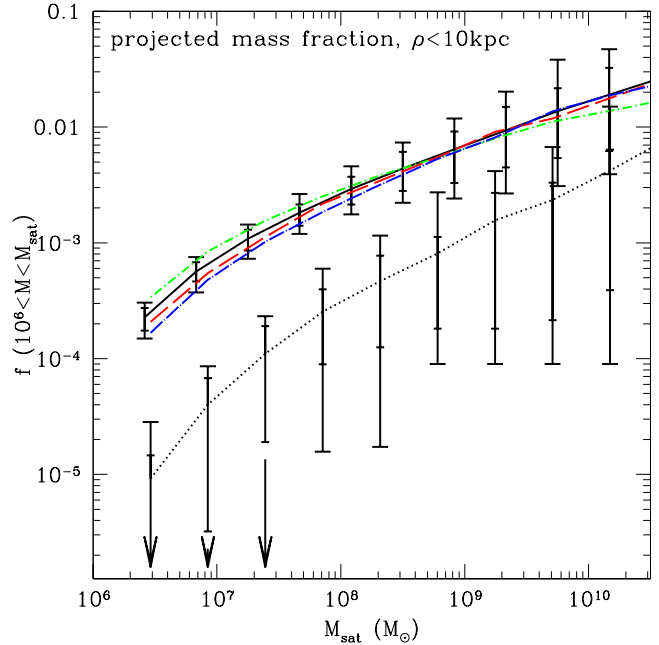}}
\caption{\label{fig:mfrac_proj}The fraction of 
mass in substructure in a central, cylindrical projection of 
radius $10$ kpc, for the same set of halos as shown in Figure \ref{fig:mfrac_total}.
The line types are the same as those in Figure \ref{fig:mfrac_total},
and again represent the mean fraction computed over all realizations.
The large and small error bars  represent the
$90$ and $64$ percentile ranges, respectively. A down-arrow
is plotted instead of a lower, large error tick if
at least $5 \%$ of the realizations had $f=0$ in that
bin. A down-arrow with no accompanying lower error bar means that
at least $18 \%$ of the realizations were without projected
substructure in that bin.}
\end{center}
\end{figure}
%
%
%
%
%
%
\begin{figure}[t]
\begin{center}
\resizebox{!}{9cm}
{\includegraphics{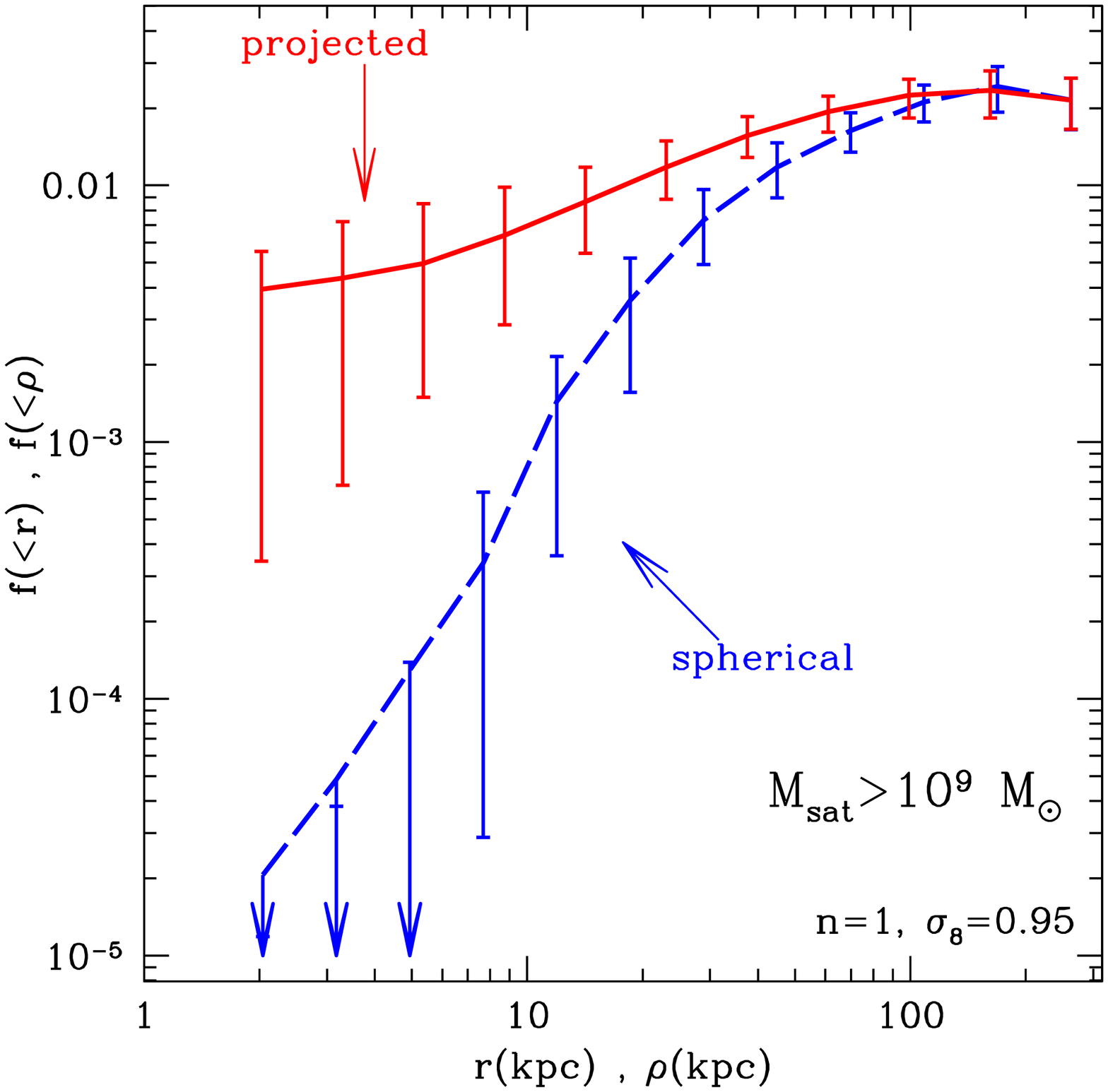}}
\caption{\label{fig:mf.comp}  The cumulative mass fraction
of substructure with $10^6$ \Msun $< M_{\rm sat} < 10^{9}$ \Msun shown 
spherically averaged as a function of radius $r$ (solid line)
and in
projection, as a function of the projected radius $\rho$ (dashed line).
The averages (lines) and $64\%$ percent ranges (error-bars) were determined
using $200$ realizations of $M=3\times10^{12}$ \Msun halos
 $z=0.6$ for our $n=1$ model.  Down-arrows indicate that more
than $18\%$ of the realizations had $f=0$ in the corresponding
radial bin.  }
\end{center}
\end{figure}
%
%
%
%
%

This  effect is illustrated  explicitly in  Figure \ref{fig:fixedvmax} 
where, rather than fixing  the host {\em mass} at $z=0$,  we have
fixed its {\em maximum  circular velocity} at $\Vmax =  187 \kms$, the 
value of a typical $n=1$, $\Mhost = 1.4  \times 10^{12}$ M$_{\odot}$
halo at  $z=0$.  Normalizing our host halos by $\Vmax$ rather than 
mass is perhaps a more reasonable choice because $\Vmax$ is more 
closely related to observations.\footnote{$\Vmax = 187 \kms$  is
somewhat smaller  than a typical  rotation velocity for a  galaxy like
the Milky Way  ($\Vmax^{\rm MW} \sim 220 \kms$), but this value  is in line
with  expectations for  the  dark  matter halo,  once  the effects  of
baryonic in-fall have been included (\eg Klypin, Zhao, \& Somerville 2002).}
Models with less galactic-scale power require a more  massive host in
order  to  obtain  the  same  value of  $\Vmax$,  and  their  velocity
functions shift  correspondingly.  For example,  a host with  $\Vmax =
187  \kms$  in  the $\sig8  =  0.65$  model requires $\Mhost \approx 
2.2  \times 10^{12}$ M$_{\odot}$.  With  this adjustment, the velocity 
functions  of the various tilted  models are now very similar.  
Again, the  BSI case is
different from  the tilted  models because the relative shift in the
$\Vmax$-$M_{\rm vir}$ relation changes abruptly with mass scale.  
It is also encouraging
that our model BSI velocity function agrees well with the
N-body results of Col{\'{\i}}n et al. (2000) for a similar type of 
truncated power spectrum (see their $R_f = 0.1$ Mpc model, Fig. 2).

Another convenient way to quantify the substructure content of halos is 
the fraction of mass in subhalos less  massive  than $M_{\rm sat}$, 
$f(<M_{\rm sat})$.  Figure \ref{fig:dfdm} shows  the
differential mass fraction $df/dM_{\rm sat}$, normalized relative to  the host mass
for several different host masses and redshifts in both the $n=1$ and BSI models.  
The results are approximately self-similar with respect to the host mass, and for the 
$n=1$ case can be well-represented by the analytic form 
\beq
\label{eq:fitfrac}
\frac{{\rm d}f}{{\rm d}x} = \lb \frac{x}{x_0} \rb^{-\alpha} \exp\lb-\frac{x}{x_0}
\rb, 
\eeq 
with $x \equiv M_{\rm sat}/M_{\rm host}$, $\alpha = 0.6$ and
$x_0 = 0.07  \pm 0.05$.  The quoted range in $x_0$ characterizes well
the rms scatter from  realization to  realization (not  shown).  This
function (with $x_0=0.07$) is  shown as the  set of crosses in Figure
\ref{fig:dfdm}.  As expected,  the mass
fractions  are somewhat  lower for the BSI model halos.  The other CDM-type
models all  yield differential mass  functions similar to those of 
the $n=1$ case.  While in the next section we present results for
a particular choice of host mass, the self-similarity demonstrated 
here implies that results at a fixed satellite-to-host mass 
ratio $x$, can be scaled in order to apply these results 
to any value of $M_{\rm host}$.
%
%
%

\subsection{\label{sub:mfrac}Mass Fractions and Gravitational Lensing}

DK01 used flux ratios in multiply-imaged 
quasars to constrain the substructure content of galactic halos to 
be $f = 0.006 - 0.07$ (at $90\%$ confidence) for 
$\Msat \lsim 10^8 -  10^{10}$ M$_{\odot}$.  In  their sample of lens systems, 
the lens redshifts span the range $0.31 \lsim
z_{\ell} \lsim 0.97$  with a median lens  redshift of $z_{\ell} \simeq
0.6$.  Our primary  goal in this  section is to make predictions
aimed at lensing studies.  Consequently, we present results for host 
systems at $z=0.6$, and with $\Mhost = 3 \times 10^{12}$ M$_{\odot}$, 
which was taken as a typical lens mass in Dalal \& Kochanek (2002, DK02). 

In  Figure \ref{fig:mfrac_total}, we exhibit results for  each  of our
inflation-derived power spectrum models.  
Here,  $f$($10^6$ \Msun$<M<M_{\rm sat}$)  is the cumulative
fraction of host halo mass that is bound up in substructures with
mass larger than $10^6$  \Msun and less  than  $M_{\rm sat}$.   As
expected from our discussion in \S \ref{sub:vfunc}, the mass fraction
in substructure is not a strong function of the tilt of the primordial
power spectrum, but it is sensitive to a sudden break in power at
small scales.   Specifically, the subhalo mass fraction in the BSI
model is roughly a factor of $\sim 3$ below that seen for the CDM-type
spectra in this mass  range.  The top   set of error bars  reflect the
$90$ percentile range derived using $200$ realizations for the $n=1$
model (other CDM-type models show similar scatter) and the bottom set
of errors reflect the same range determined  from $50$ realizations of
the BSI spectrum.

Rather   than the  total   mass  fraction,  lensing  measurements  are
sensitive   to the mass fraction  in   substructure projected onto the
plane of the lens at a halo-centric distance  of order the  Einstein
radius of the lens, $R_{\rm E} \sim 5-15$ kpc.  In Figure 
\ref{fig:mfrac_proj} we  plot $f_{\rm sat}$($>10^6$ M$_{\odot}$) 
projected  through a cylinder of radius $10$ kpc centered on the host
halo for the same set of halos shown in Figure \ref{fig:mfrac_total}.  
The large and  small  error bars reflect the
$90$ and  $64$ percentile ranges,  respectively, in measured projected
mass fractions derived using 200 $n=1$ realizations (top set) and  50
BSI realizations (bottom  set).  A down-arrow  is plotted instead of a
lower, large  error tick if at  least  $5 \%$ of the  realizations had
$f=0$ in that bin.  A  down-arrow with no  accompanying lower error bar
indicates that at least $18 \%$ of the realizations were without projected
substructure in that bin.

%
\begin{figure}[t]
\begin{center}
\resizebox{!}{9cm}
{\includegraphics{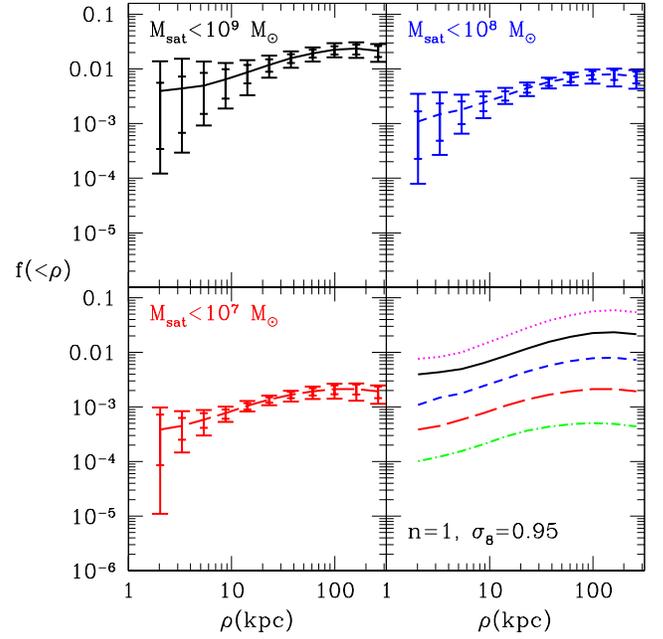}}
\caption{\label{fig:split.neq1}Mass fraction 
in substructure in cylindrical projection of 
radius $\rho$ for the same set of $n=1$ halos described
in Figure \ref{fig:mf.comp}.  The upper left, upper right, and
lower left panels show the mass fraction profiles in subhalos
larger than $10^{6}$ \Msun and less than $10^9$, $10^8$ and 
$10^7$ \Msun respectively.  Error bars reflect the same percentile
ranges as do those in Figure \ref{fig:mf.comp}.  The bottom right
panel illustrates how 
the mean projected mass fraction profiles vary as a function of
the maximum subhalo mass considered: $10^{6.3}$, $10^7$, $10^8$, 
$10^9$, and $10^{10}$ \Msun from bottom to top.}
\end{center}
\end{figure}

\begin{figure}[t!]
\begin{center}
\resizebox{!}{9cm}
{\includegraphics{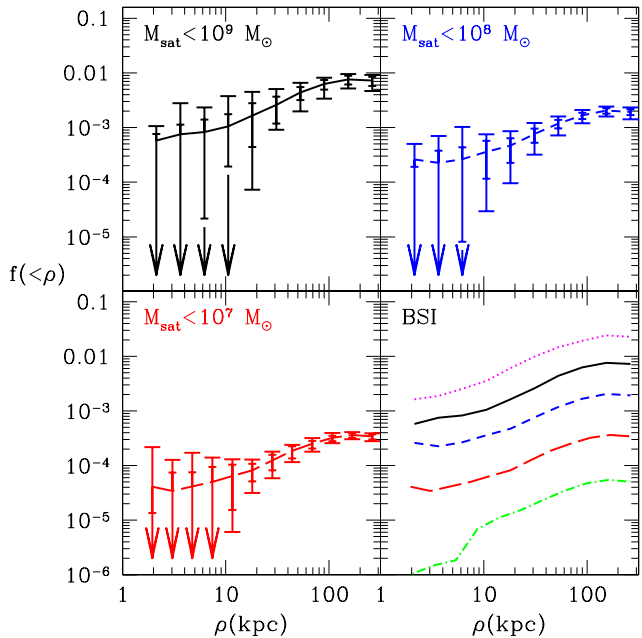}}
\caption{\label{fig:split.bsi} Mass fraction 
in substructure in cylindrical projection for the BSI model.
All line types and panels are  as described in Figure
\ref{fig:split.neq1}.}
\end{center}
\end{figure}

The projected  mass fractions are  not as severely suppressed
relative to the volume-averaged mass fractions as one might 
expect given that  tidal forces act systematically to destroy 
substructure near host halo centers (see Figure \ref{fig:radial}).  The reason
is that we are examining substructure in a cylindrical volume, and
picking up subhalos with large halo-centric radii.  We illustrate this
effect in Figure \ref{fig:mf.comp}, where we compare the mass fraction
in  cylindrical projection radius $\rho$  with  the mass fraction  in
spherical shells with the same value of spherical radius $r$.  Notice
that the mass fraction  in spherical regions is significantly  reduced
in  the center,  while the projected  mass
fraction is less severely affected.  Of course, the mass fraction
approaches the global value at large radii.  Figures
\ref{fig:split.neq1}  and \ref{fig:split.bsi} demonstrate how the mass
fractions change as a function of projection radius for various
subhalo mass cuts for the $n=1$ and BSI models respectively.
Notice that the relative drop in mass fraction as
a function of projection radius is more pronounced in the BSI model
than in the $n=1$ case.  This reflects the fact that tidal disruption 
is more important in the BSI case and core-like behavior of the subhalo 
radial distribution sets in at a larger radius in this model.  
%
%
%
%

\subsection {Warm Dark Matter and Gravitational Lensing}

In the previous section  we demonstrated  that the  substructure mass
fraction is sensitive to abrupt changes in the power spectrum and
used the BSI model as a specific example. 
In this section we investigate these differences in the context of 
WDM.  We label the different WDM models by the warm particle mass $m_{\rm W}$,  
and assume the canonical case of a ``neutrino-like'' thermal relic 
with two internal degrees of freedom, $g_{\rm W}=2$.

\begin{figure}[t!]
\begin{center}
\resizebox{!}{9cm}
{\includegraphics{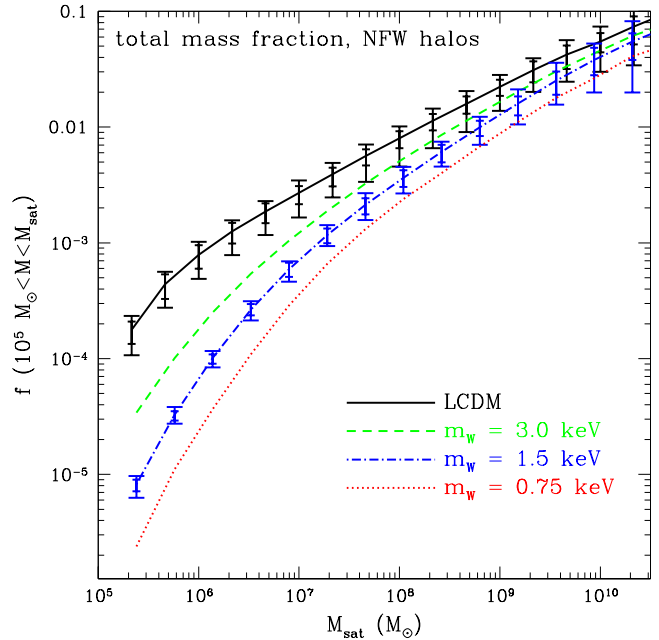}}
\caption{\label{fig:mfrac_wdm1} Total cumulative mass fractions 
in substructure more massive than $10^5$ \Msun for the
$n=1$ and 3 WDM models.  The models are $\Lambda$CDM (solid), 
$m_{\rm W}=3.0$ keV (dashed), $m_{\rm W}=1.5$ keV (dash-dot), and 
$m_{\rm W}=0.75$ keV (dotted).  For clarity, we show error bars only 
for the $\Lambda$CDM and $m_{\rm W}=1.5$ keV models.  The error bars 
and arrows have the same meaning as in Figure \ref{fig:mfrac_proj}.}
\end{center}
\end{figure}

\begin{figure}[t!]
\begin{center}
\resizebox{!}{9cm}
{\includegraphics{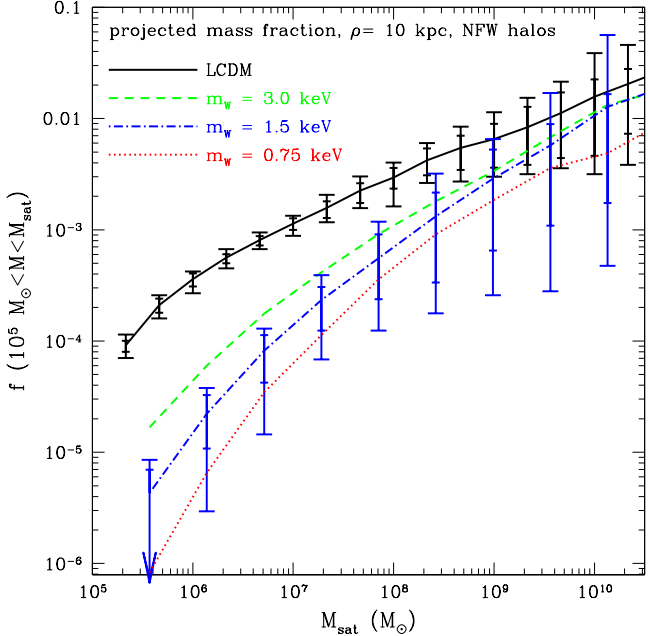}}
\caption{\label{fig:mfrac_wdm2} Cumulative mass fractions 
in substructure more massive than $10^5$ \Msun within a projected radius 
of $10$ kpc for the same models shown in Figure \ref{fig:mfrac_wdm1}.  
The linetypes are the same as in Figure \ref{fig:mfrac_wdm1}.}
\end{center}
\end{figure}

Figure \ref{fig:mfrac_wdm1} shows the total mass fraction of 
$3 \times 10^{12}$ M$_{\odot}$ host halos at $z=0.6$ as a
function of $M_{\rm sat}$ implied by our three WDM model power spectra
compared to  our standard $\Lambda$CDM case.  For substructure  smaller than 
$\sim 10^{7}$ M$_{\odot}$, the differences between the models are as large as 
an order of magnitude or more, and even the largest WDM particle mass ($3$ keV) 
provides a potentially measurable suppression of substructure.  Figure 
\ref{fig:mfrac_wdm2}  shows  the mass fraction in projected 
cylinders of radius $10$ kpc.

The differences  in mass  fractions seen for  the different  models in
Figures  \ref{fig:mfrac_wdm1}   and  \ref{fig:mfrac_wdm2}  come  about
because subhalos become less concentrated relative to their host halos
as the WDM particle mass is decreased and power is suppressed on larger
scales, much like the BSI case.  In true WDM
models there  are additional processes  that, in principle, can  alter the
formation and density structure of dark matter halos.  In Figures 
\ref{fig:mfrac_wdm1} and \ref{fig:mfrac_wdm2}, we have only accounted 
for the effect of the power spectrum on substructure mass fractions and 
assumed that the density structure of WDM halos is {\em identical} to 
that for CDM halos.  For high mass systems, this is a sensible approximation  
(Col{\'{\i}}n et al. 2000a; Avila-Reese et al. 2001), but this 
approximation should break down at small masses and lead to further 
suppression of substructure.  

One  consequence of a WDM particle with non-negligible velocity
dispersion is that gravitational clustering is resisted by structures
below the effective Jeans mass of the warm particles 
(\eg Hogan \& Dalcanton 2000; Bode et al. 2001):
\beq
\label{eq:Jeans}
M_{\rm J} \approx 6 \times 10^3 \Bigg(\frac{{\rm keV}}{m_{\rm W}}\Bigg)^4
\Bigg(\frac{\Owdm h^2}{0.15}\Bigg)^{1/2}\Bigg(\frac{2}{g_{\rm W}}\Bigg) 
(1+z)^{3/2} M_{\odot}.
\eeq
For both the $m_{\rm W} =  1.5$ keV and $m_{\rm W} = 3.0$ keV models, 
$M_{\rm J} \ll 10^5 M_{\odot}$ when $z \lsim 10$, so all halos  of 
interest in this context are minimally affected.  The situation is 
somewhat more complicated in the  $m_{\rm W} = 0.75$  keV model, 
where $M_{\rm J} \gsim 10^5$ M$_{\odot}$ for
redshifts  $z \gsim 2$.  We therefore expect that the formation
of these halos should be suppressed compared to the predictions of the
EPS formalism.  This suppression should only have a minor effect on
our predictions because we restrict ourselves to satellite  masses
$\gsim 10^{5}$ \Msun and most surviving subhalos are accreted at $z \lsim 2$.  
In the interest of simplicity, we chose to ignore this effect here.  
As a result, we may significantly {\em over-predict} substructure mass 
fractions at low $M_{\rm sat}$ in these cases.  In the context of this 
study, this is a conservative approach because the true mass fraction 
would be reduced by these effects, bringing it further away from the 
measured substructure mass fractions and standard $\Lambda$CDM predictions.

%

In addition to the effective Jeans suppression, WDM halos, unlike their 
CDM counterparts, cannot achieve extremely high densities in their centers
due to phase space constraints (Tremaine \& Gunn 1979).  In the early 
Universe the  primordial phase space distribution of the WDM particles 
is a Fermi-Dirac distribution with a maximum of 
$f_{\rm  max} = g_{\rm W}/h_{\rm  PL}^3$ at low  energies 
($h_{\rm PL}$ is Planck's constant).  For a collisionless species, the
phase space density is conserved and this maximum phase space density
may not be exceeded within WDM halos.  If we define the phase density
as $Q \equiv \rho/(2\pi\sigma^2)^{3/2}$ then the maximum allowed phase 
density is (Hogan \& Dalcanton 2000) 

\beq 
\label{eq:Qmax}
Q_{\rm max} \simeq 5.2 \times 10^{-4} 
\Bigg(\frac{m_{\rm W}}{{\rm keV}}\Bigg)^4 \Bigg(\frac{g_{\rm W}}{2}\Bigg) \ 
\frac{{\rm M}_{\odot}/{\rm pc}^3}{({\rm km}/{\rm s})^3}.
\eeq
This limit implies that WDM halos cannot achieve the central 
density cusps of the kind observed in simulated CDM halos.   Instead, we 
expect a core in the density profile.  For viable WDM models,  the phase
space  core is  expected to  be dynamically  unimportant for any halo  
massive enough to host a visible galaxy (ABW).  However, for the 
lowest-mass subhalos ($M \lsim 10^7$ M$_{\odot}$) the presence of 
phase space-limited cores may be important because halos with large 
cores are less resistant to tidal forces than cuspy halos.


%
%
We have  attempted to estimate (crudely) how the  phase space
limit affects the substructure population of WDM halos by adopting
our  standard  model of  halo  accretion  and  orbital evolution,  but
allowing the density structure  of the appropriately small subhalos to
be set  by the phase  space limit.  For  these calculations we  used the
phenomenological density profile of Burkert (1995), 

\beq
\label{eq:burk_profile}
\rho_{\rm B}(r) = \frac{\rho_0}{[1+r/r_{\rm B}][1+(r/r_{\rm B})^2]}.
\eeq 
The Burkert profile resembles the NFW form at large radius, but 
features a constant density core at its center, and thus a 
velocity dispersion that approaches a constant at small $r$:
$\sigma_0^B  \simeq 0.55 \Vmax$.\footnote{The numerical coefficient in 
Eq. (17) of ABW should be $\approx 0.3$ rather than $0.2$.}  For Burkert profiles, 
$\Vmax^2 \simeq 0.86\Vvir^2 c_{\rm B}/g_{\rm B}(c_{\rm B})$, where 
$c_{\rm B} \equiv R_{\rm vir}/r_{\rm B}$ is the Burkert concentration and 
$g_{\rm B}(y) \equiv \ln (1+y^2) + 2\ln (1+y)-2 \tan^{-1}(y)$.  
Solving for the phase density in the core ($r \ll r_{\rm B}$) and equating 
it with the maximum phase density of equation (\ref{eq:Qmax}) yields the 
following relation for the maximum attainable value of $c_{\rm B}$: 

\begin{eqnarray}
\label{eq:c_B_max}
c_{\rm B}^{3/2}g_{\rm B}^{1/2}(c_{\rm B}) & \simeq & \frac{111}{(1+z)^3} \Bigg(\frac{0.15}{\Ow h^2}\Bigg) 
\Bigg(\frac{178}{\dvir}\Bigg) \Bigg(\frac{\Vvir}{{\rm km/s}}\Bigg)^3 \nonumber \\
 & & \times \Bigg(\frac{g_{\rm W}}{2}\Bigg) \Bigg(\frac{m_{\rm W}}{{\rm keV}}\Bigg).
\end{eqnarray}

We assigned Burkert  concentrations  according to  the following
prescription.  First, we computed NFW concentrations $\cvir$, for each
halo according to the B01 model.   We converted from NFW concentration
to Burkert concentration $c_{\rm B}$, by interpreting the B01 value of
$r_{\rm s}$ as  the radius at which $\dd \ln \rho(r)/ \dd \ln  r
|_{r=r_{\rm s}} = -2$.  This implies that $r_{\rm B} \simeq 0.66
r_{\rm s}$   or $c_{\rm  B} \simeq 1.5 c_{\rm vir}$.  With this
correspondence, the adopted Burkert profile achieves the maximum of
its  rotation curve at $r_{\rm max}^{\rm  B} \simeq 0.99 r_{\rm max}$,
where $r_{\rm max}$ is the radius  at which the corresponding NFW halo
achieves $\Vmax$.   Similarly, $\Vmax$ of the adopted Burkert profile is 
within  $10\%$ of the corresponding NFW $\Vmax$ for all relevant 
concentrations  ($1\le  \cvir \le   25$).
Second, we  computed the maximum value  of $c_{\rm  B}$ allowed by the
phase space constraints using Eq. (\ref{eq:c_B_max}).  We  then
assigned each halo the smaller of these two values of $c_{\rm B}$ at the 
time of accretion.  In this way, we guaranteed that the phase space 
constraint was  met by all halos.  We have checked that this  prescription for 
Burkert halos does not yield any systematic bias in our results by applying it 
all of our CDM models.  We found that it gave nearly identical results to that 
of our standard NFW model, which is not  surprising  in the context of  our
model and disruption criteria.

\begin{figure}[t!]
\begin{center}
\resizebox{!}{9cm}
{\includegraphics{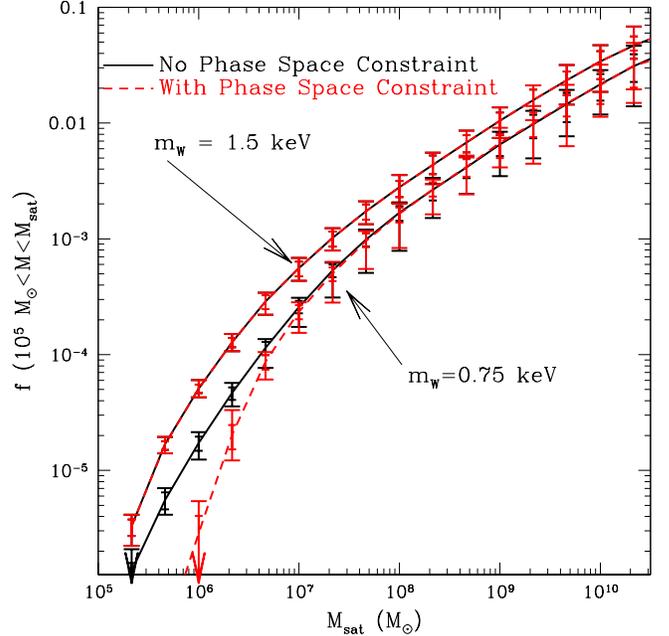}}
\caption{\label{fig:mfrac_wdm3} Total cumulative mass fractions 
in substructure more massive than $10^5$ \Msun both with
(dashed) and without (solid) estimating the effects of the
phase space limit.  The upper set of lines corresponds to the 
$m_{\rm W}=1.5$ keV model and the lower set of lines corresponds to 
the $m_{\rm W}=0.75$ keV model.  The error bars and arrows are as in 
Figure \ref{fig:mfrac_proj}.
}
\end{center}
\end{figure}

We present our estimates of cumulative mass fractions in WDM models, including 
the effect of the phase space constraint, in Figure \ref{fig:mfrac_wdm3}.    
It is clear, at least from this rough estimate, that the
Tremaine-Gunn limit plays an important role only  for the most extreme
WDM models, $m_{\rm W} \lsim 1$ keV, and only the smallest halos, $\lsim
10^6$ M$_{\odot}$.  However, we emphasize that our new 
assumptions about WDM halos have {\em not} been tested with N-body simulations.    
Simulations have yet to examine the density structure of
halos that saturate the phase space bound and most studies 
have ignored the initial velocity  dispersion of the WDM particles
(Col{\'{\i}}n et al. 2000a;  Avila-Reese    et al.  2001, Knebe  et
al. 2002), but the Burkert profile assumption seems plausible.  With
these precautions in mind, Figure \ref{fig:mfrac_wdm1} may be regarded
as an approximate upper-limit on the substructure mass fraction for WDM halos.
Any  phase space bound  or the   effects  of   primordial  velocity
dispersions on halo  formation and density  structure should lead
to enhanced disruption, resulting in lower mass fractions.

%

%

One  physical process that might affect WDM (and BSI) models that we
have not considered is top-down fragmentation (\eg Knebe et al. 2002).
It is possible  that  power can be transported from large  scales to
small in truncated models, resulting in a population of low-mass halos
that would not be accounted for in Press-Schechter theory.  While such
a process  could result in a higher substructure abundance  than that
estimated using our model, there are  reasons to believe that
the effect should be fairly small.  Systems  that form in
this manner collapse quite late, and  their density structure
likely would be very diffuse compared to  their hierarchically-formed 
brethren.  Therefore, it is less likely that systems formed via
fragmentation could survive tidal disruption once incorporated into a
galactic halo.  

%
%
%

\subsection{\label{sub:dwarf}The Dwarf Satellite Problem}

Comparisons between the predicted subhalo population  and the observed
dwarf galaxy abundance are usually made by comparing  counts as  a
function of maximum circular velocity, $\Vmax$.  
This is a sensible mode of comparison because it sidesteps the 
complicated issues of star formation and feedback in these 
poorly-understood galaxies.  Yet, there  are considerable uncertainties, 
even for this method of comparison, and it is likely that efforts to 
compare predictions as a function of dwarf luminosity 
(Somerville 2002; Benson et al. 2002) in tandem with velocity 
comparisons will be needed in order to fully understand the nature of 
the dwarf satellite problem.

For most satellites, the quantity that is observed and used to infer 
the halo $\Vmax$ is the line-of-sight stellar velocity dispersion, 
$\sigma_{\star}$.  As discussed in S02 and H03, the mapping between 
$\svd$ and $\Vmax$ depends upon the theoretical expectation for the 
density profile  of the subhalo as well as on the stellar mass 
distribution of the galaxy.  An additional complication concerns 
the unknown velocity anisotropy of the stars in the system.

\begin{figure}[t]
\begin{center}
\resizebox{!}{9cm}
{\includegraphics{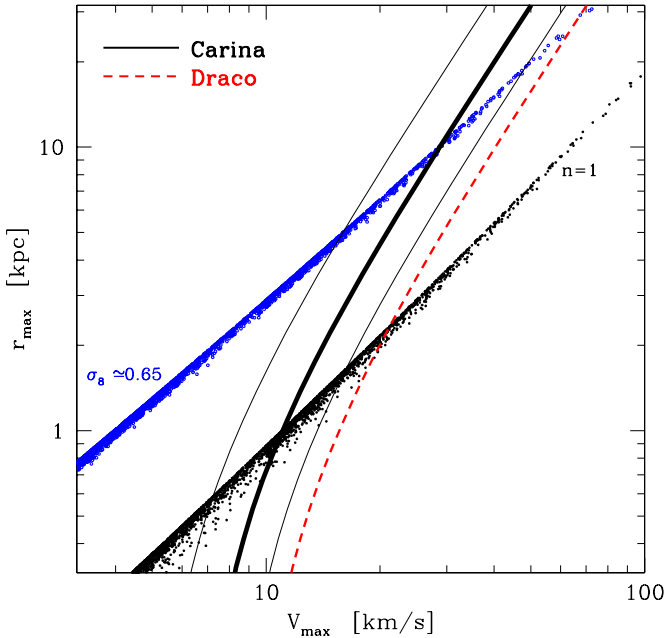}}
\caption{\label{fig:svd} The lower and upper sets of points show  a scatter 
plot of $\Vmax$ and $\Rmax$ for all surviving halos produced in 10 merger 
tree realizations for the $n=1$, and low-normalization, $\sig8=0.65$ models 
respectively.  The thick solid line shows the locus of points in 
the $\Vmax$-$\Rmax$ plane that corresponds to the central value of the measured velocity 
dispersion of Carina ($\svd =6.8 \pm 1.6 \kms$) given the measured King 
profile parameters of Carina (Table \ref{table:vmax}) and assuming NFW halos.  
The thin solid lines correspond to the $\pm 1\sigma$ quoted errors in 
velocity dispersion for Carina.  The dashed line corresponds to the 
locus of points that is consistent with the central value of the 
measured velocity dispersion of Draco ($\svd = 9.5 \kms$).}
%
\end{center}
\end{figure}

A phenomenologically-motivated approximation for the stellar distribution
in a dwarf galaxy is the spherically symmetric King profile (King 1962),
\beq
\label{eq:king}
\rho_{\star}(r) = \frac{k}{z^2} \Bigg(\frac{\cos^{-1}(z)}{z} - \sqrt{1-z^2}\Bigg), 
\eeq
where $z \equiv [1 + (r/r_{\rm c})^2]/[1 + (r_{\rm t}/r_{\rm c})^2]$,  
$r_{\rm c}$ and $r_{\rm t}$ are the core and tidal radii of the King profile, 
and $\rho_{\star}(r>r_{\rm t}) = 0$.  The normalization is not important in 
what follows. 

If we assume that a stellar system described by Equation (\ref{eq:king}) is 
in equilibrium and embedded in a spherically symmetric dark matter potential 
characterized by the circular velocity profile $V_{\rm c}(r)$, then the radial 
stellar velocity dispersion profile $\sigma_{\rm r}(r)$, can be computed via the 
Jeans equation:
\beq
\label{eq:jeans}
r \frac{\dd [\rho_{\star} \sigma_{\rm r}^2]}{\dd r} = - \rho_{\star}(r) V_c^2(r)
        - 2 \beta(r) \rho_{\star}(r) \sigma_{\rm r}^2(r),
\eeq
where the anisotropy parameter $\beta \equiv 1 - \sigma_{\bot}^2/2\sigma_r^2$, 
$\sigma_{\bot}$ is the tangential velocity dispersion, and $\beta=0$ corresponds 
to an isotropic dispersion tensor.  A measured, line-of-sight velocity dispersion
is determined by the projected velocity dispersion profile weighted by the luminosity 
distribution sampled along the line-of-sight.  For the
isotropic case it is given by 
\beq
\label{eq:sigstar}
\sigma_{\star}^2 = \frac{\int_{0}^{r_{\rm t}}\rho_{\star}(r')V_{\rm c}^2(r')
\textrm{d}r'}{\int_{0}^{r_{\rm t}}\rho_{\star}(r')\textrm{d}r'}, 
\eeq
assuming a constant mass-to-light ratio.
If a galaxy has a measured stellar profile (the King parameters in this case) 
and measured value of $\sigma_{\star}$, then the Jeans equation 
places only one constraint on the rotation curve of the system, $V_{\rm c}(r)$.
We expect the halo velocity profile to be at least
a two-parameter function (\eg the NFW profile) so determining $\Vmax$ 
requires some theoretical input for the expected form of $V_{\rm c}(r)$ 
in order to provide a second constraint.

Motivated by dark matter models, we assume  that the global 
rotation curve is set  by an NFW profile associated with the 
dwarf galaxy halo.  The rotation curve for an NFW halo
is fully described by specifying two parameters and a natural pair 
is $\Vmax$ and $\Rmax$.  For any given
cosmology, the relation between $\Vmax$  and $\Rmax$ is expected to be
rather  tight, and this provides a second, theoretically-motivated 
constraint that sets the $\Vmax$-$\svd$ mapping implied by Eq. (\ref{eq:jeans}).

The  $\Vmax$-$\Rmax$ relationships for surviving subhalos in 
two of our models are shown in Figure \ref{fig:svd}.  The 
lower set of points corresponds to our standard $n=1$ model 
and the higher set of points is derived from our $\sig8 = 0.65$ model.
In each case, we plot one point for each surviving halo in 10 model realizations.
The strong correlation, $\Rmax \propto \Vmax^\gamma, \gamma \simeq 1.3$, 
follows directly from the input correlations between 
$\Mvir(z)$, and $\cvir$ (see \S \ref{sub:desc} and B01).  
The normalizations and slopes are influenced by the cosmology, 
accretion times and (mildly)
by the orbital history of the subhalos.\footnote{The 
scatter in the $\Vmax$-$\Rmax$ plane should be larger
than that shown here
because we have not included the expected scatter in the input 
$\cvir$-$\Mvir$ relation.  For $\sigma(\log\cvir) \approx 0.14$ 
(B01; W02), the implied scatter is $\sigma(\log\Rmax) \approx 0.18$ at fixed $\Vmax$.  
For $\sigma(\log\cvir) \approx 0.08$ (Jing 2000), the implication is
$\sigma(\log\Rmax) \approx 0.11$.}  

The thick solid and dashed lines in Figure \ref{fig:svd} 
show the locus of points in the $\Vmax$-$\Rmax$ plane that 
correspond to the central values of the observed line-of-sight velocity dispersions 
for Carina and Draco respectively, given their measured King profile parameters.    
Our adopted $\svd$ values  and King parameters
are listed in Table \ref{table:vmax} along with appropriate references.
The  light solid lines  illustrate  how these contours  expand  when we
include  the $\pm 1\sigma$ measurement error in $\svd$ for Carina.  A
similar (although narrower) band exists for Draco, but we have omitted it 
for the sake of  clarity. Consistency with the observed King parameters 
and velocity dispersions requires each dwarf to reside in a halo with 
structural parameters that lie within the region of overlap between the 
contours and the model points.  For example, in the $n=1$ model Carina is expected
to reside in a halo with $\Vmax \approx 11  \kms$ and  $\Rmax  \approx 1$ kpc.  
For the $\sig8 = 0.65$ model, Carina is expected to sit in a {\em larger} halo,
with $\Vmax \approx 29 \kms$ and  $\Rmax \approx 10$ kpc. 
Similar comparisons hold for Draco and all of the 
Local Group dwarf satellites and these comparisons can be made 
in a similar way for any cosmology.
The point is that the maximum velocities that are assigned to 
satellite galaxies are cosmology-dependent.  Therefore, ``observed'' 
velocity functions are also cosmology-dependent because theoretical 
inputs are used to convert from $\svd$ to $\Vmax$.

In Table \ref{table:vmax} we show estimates  for halo $\Vmax$ values for the 
observed Milky Way satellites under the assumption
that $\beta = 0$ along with a similar analysis for 
$\beta = 0.15$ (values in parentheses).  
We estimate halo $\Vmax$ values for six
different power spectra, relying on the model-dependent
$\Rmax$-$\Vmax$ relationship for substructure in each case, and taking the 
central values of the measured velocity dispersions for each halo.  Taking 
the quoted $\pm 1\sigma$ range for the measured velocity dispersions typically 
leads to a shift in $\Vmax$ of $\sim 30 \%$ which is considerable compared 
to the inherent scatter in the $\Mvir-\cvir$ relation.  For
reference, we  have also included  the adopted $\Vmax$ values from the
original K99 work on the  dwarf satellite
problem.  As expected, the implied $\Vmax$  values become larger as we
explore models with less  galactic-scale power.  Our estimates
for the $n=1$, $\beta = 0$ case are close to those of K99.

\begin{figure*}[t]
\centerline{ 
   \epsfysize=3.5truein  \epsffile{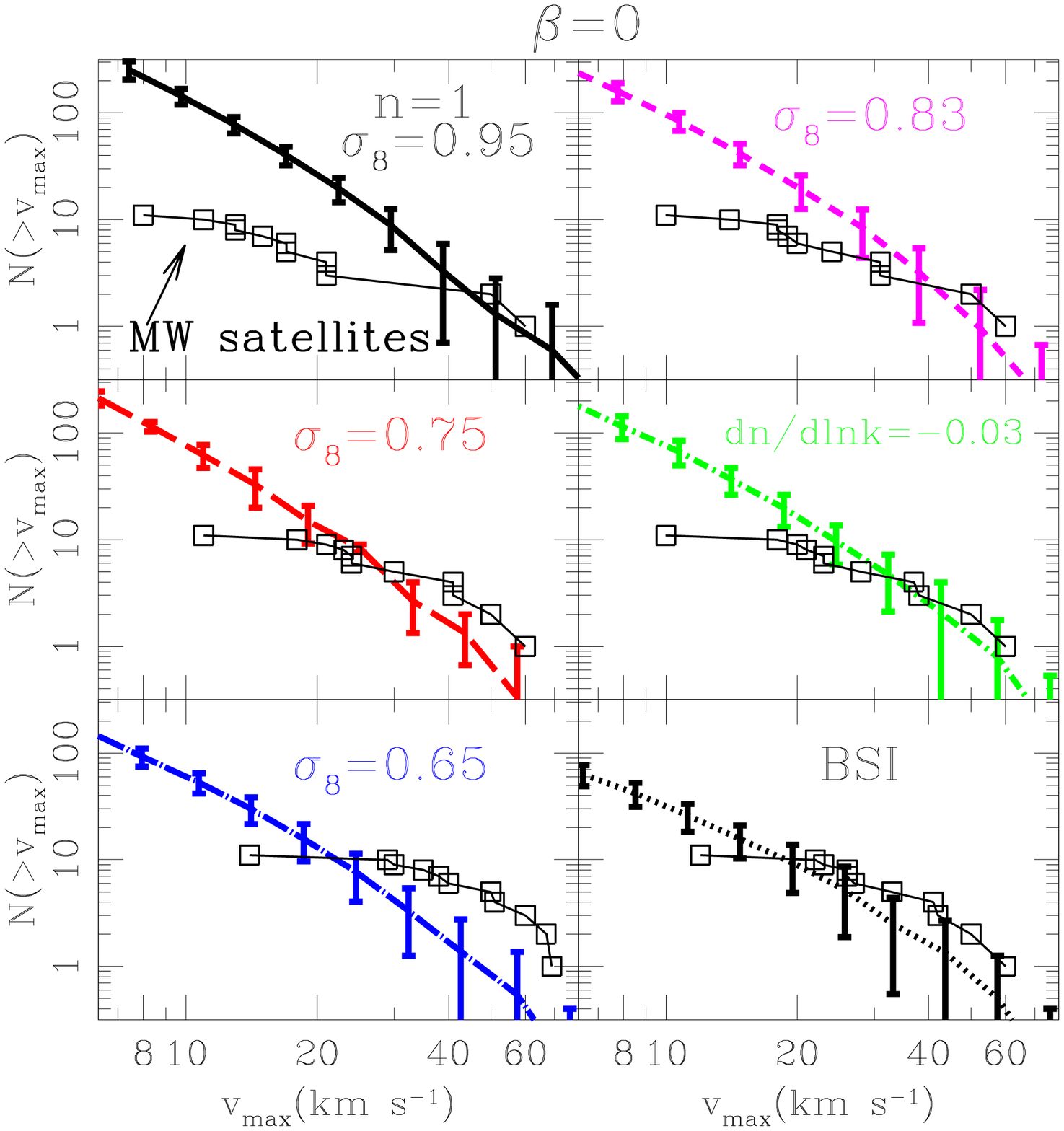}
   \epsfysize=3.5truein  \epsffile{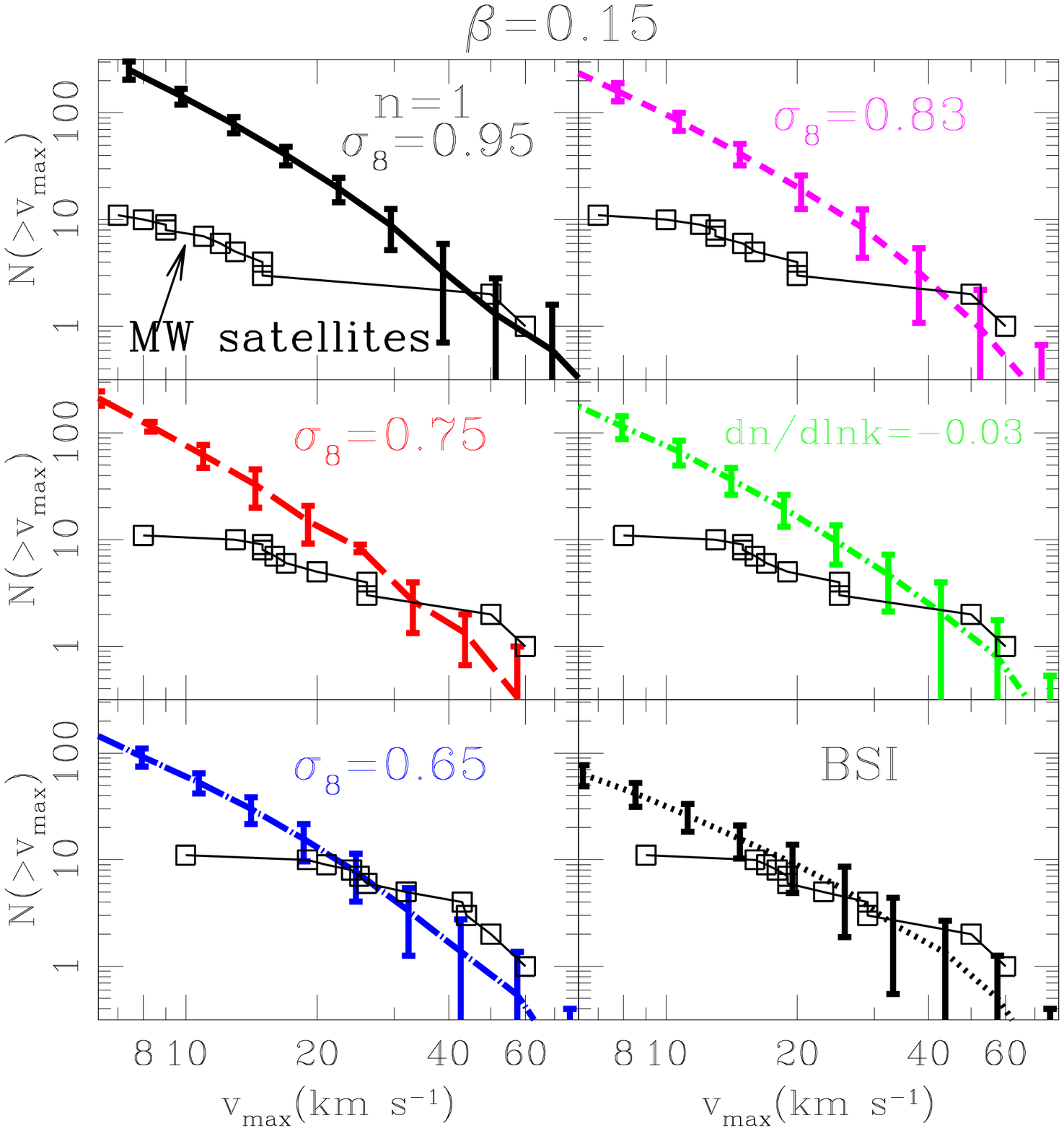}
}                                                  
\caption{\label{fig:vfunc_comp}Satellite halo 
velocity functions for six of our models
compared with the velocity functions of the Milky Way 
satellites after accounting for the cosmology-dependent 
mapping between $\svd$ and $\Vmax$.  
The squares represent the velocity function of 
Milky Way satellites based on the data in Table \ref{table:vmax}.  
The lines represent the means over all realizations and the error 
bars reflect the dispersion among these realizations.  The models 
are labeled in each panel.}
\end{figure*}

The left panel of Figure  \ref{fig:vfunc_comp} shows the Milky Way 
satellite counts for each model, assuming $\beta = 0$, along with 
the predicted velocity functions for each model.  In addition to the 
satellites listed in upper portion of Table \ref{table:vmax}, 
we have  also included the  Small Magellanic Cloud (SMC, 
with $\Vmax  = 60  \kms$,  estimated by
Stanimirovi{\'c} 2000 to include a substantial contribution from the
baryonic component) and the Large Magellanic Cloud (LMC, 
with $\Vmax = 50 \kms$, van der Marel
et al.  2002) in our cumulative velocity  functions.  In the standard
case  ($n=1$) the  discrepancy  sets   in at   $\Vmax \sim  30 \kms$,
requiring roughly one-in-ten halos with $\Vmax \sim 10-20 \kms$ to be the host of 
an observed galaxy.  The extremely tilted model with $\sigma_8 \simeq 0.65$ actually
under-predicts the dwarf count for large  systems.   Interestingly,
dwarfs  in  the ${\rm d}n/{\rm d}\ln k = -0.03$ RI model as well as 
the $\sigma_8=0.75$  case are consistent with inhabiting the ten  
most massive subhalos, with only Sextans standing as an outlier.  
The BSI  model also looks to  be in good agreement with the data for $\Vmax
\gsim  12   \kms$, lending support to the conclusions of 
Kamionkowski \& Liddle  (2000).  However, the problem of Local Group 
satellites is not completely ``solved'' in any of these models  because the
velocity function continues to  rise below the velocity-scale of
Sextans in all cases.   What changes in the low-power models is the
nature of  the discrepancy.  In one extreme, the mismatch  sets in at
$\Vmax  \sim  30  \kms$   and  gradually  becomes  worse  for  smaller
systems.  In the other extreme, the  mismatch  seems to  imply a 
sharp threshold for dwarf galaxy formation at a scale near 
$\Vmax \sim 10-20 \kms$. 

Unfortunately, a  detailed accounting of the mismatch  is difficult,
even  for  a given  cosmology.  The  dwarf  $\Vmax$
estimate  is  very sensitive  to  the  velocity anisotropy  parameter,
$\beta$.  For example, assuming $\beta = 0.15$ leads to $\Vmax$ values that 
are significantly lower than in the isotropic case because rotational support 
has been traded for pressure support.  The velocity function comparisons with $\beta=0.15$ are  
shown in the right panel of Figure \ref{fig:vfunc_comp} 
and the $\Vmax$ values are listed in parentheses 
in Table \ref{table:vmax}.  In this case, only the $\sigma_8 \simeq 0.65$ 
case and the BSI  model can account for the dwarf population without 
a differential feedback mechanism.
The rest of the models over-predict the counts, with the greatest apparent 
discord in the $n=1$ case.  The  specific choice of $\beta  = 0.15$
serves mainly to illustrate the effect of a minor anisotropy.  We 
chose this value because it is typical of what is seen in the central
regions of simulated dark matter halos (Col{\'{\i}}n et al. 2000b), 
and therefore it seems a reasonable possibility for the 
anisotropy parameter of particles in dwarf galaxies.

%
%
\section{\label{sec:caveats}Caveats}

In this study we employed a simple, semi-analytic
model based on many previous studies (LC93; SK99; BKW; TB01; B01; W02; HFM02) 
and designed to produce large  numbers  of halo realizations with minimal
computational effort.  In   developing this model,  we  have made many
simplifying assumptions.  In this section we draw attention to many of
these shortcomings and discuss how they might  affect our results and
be improved upon in future work.

Among the most obvious omissions in this work is the neglect of any
disk or bulge component in each  halo.  We have specifically chosen to
ignore  the  effects of central galaxies  because  the physics of dark
halo formation is   relatively well-understood compared  with that  of
galaxy formation.  This allows us  to ground  our  work against
dissipationless N-body simulations.   In  order to include  a galactic
component,  one is forced to adopt 
many poorly-constrained models and assumptions regarding
 gas accretion,  cooling,  angular   momentum
distributions, feedback, and  the  effects of substructure on
the host galaxy itself.  Once a reliable  framework for the dark matter has
been developed, we  can use this as  a foundation for more speculative
(yet interesting)     explorations  involving the     baryonic
components.

A  central  (disk) galaxy would add   to the  dynamical friction force
experienced by  subhalos orbiting near the plane  of  the disk and cause
halos on highly inclined orbits to be tidally heated during 
rapid encounters with the disk potential 
(\eg Gnedin \&   Ostriker   1999; Gnedin, Hernquist, \&
Ostriker 1999; TB01).  These effects lead to enhanced satellite 
disruption.  Conversely, subsystems that are massive
enough to host galaxies might be  rendered more resistant to tidal
disruption because their central densities would be enhanced by the 
presence of cool baryons.  For low-mass halos, the net effect of a 
central galaxy would likely be to reduce the substructure count,  
mainly at small radii.  Even without  including  these effects, we find  that
the substructure fraction  drops significantly at small radii because
of the dark  matter potential, and that a  large part of the projected
central  mass fraction comes  from  subhalos at  large  radii that are
picked up in projection.   Nevertheless, projected mass  fractions are
rather sensitive   to  the size  of the   core  in the subhalo  radial
distribution (Chen et al.  2003), so if the core region were larger 
as a result of a central galaxy, the implied lensing signal would 
be reduced relative to our estimates.  We find that eliminating 
{\em all} substructure within $20$ kpc of the halo center, reduces 
the projected mass fractions in subhalos less massive than 
$10^{10}$ M$_{\odot}$, $f_{10}$, by $\sim 30\%$.  

\onecolumn

\newpage

\begin{deluxetable}{llccccccccc}
\tablewidth{7.5in}
\tablecaption{\label{table:vmax}Characteristics of
Milky Way satellites. }
\tablehead{
\colhead{Satellite} & \colhead{$\svd$} & \colhead{$r_{c}$} 
& \colhead{$r_{t}$}  &  \colhead{$\Vmax$} 
& \colhead{$\Vmax$} & \colhead{$\Vmax$} & \colhead{$\Vmax$} 
& \colhead{$\Vmax$} & \colhead{$\Vmax$} & \colhead{$\Vmax$} \\
 & &  &  & \colhead{K99} & \colhead{$n=1$}  & \colhead{$\sigma_8=0.83$} 
 & \colhead{RI model} & \colhead{$\sigma_8=.75$} & \colhead{BSI} & \colhead{$\sig8=0.65$} \\
 & \colhead{($\kms$)} & \colhead{(kpc)} & \colhead{(kpc)} & \colhead{($\kms$)} & \colhead{($\kms$)} & 
\colhead{($\kms$)} & \colhead{($\kms$)} & \colhead{($\kms$)} & 
\colhead{($\kms$)} & \colhead{($\kms$)}
}
\startdata
Sagittarius & $11.7 \pm 0.7$ & 0.44 & 3.0 & 20 &  17 (13)  & 20 (15)  & 23 (17) & 24 (17) & 26 (19) & 35 (25)  \\
Fornax      & $10.5 \pm 1.5$ & 0.46 & 2.3 & 18 &  15 (11)  & 18 (13) & 20 (15) & 21 (15)  & 23 (17)  & 30 (21)  \\
Draco       & $9.5 \pm 1.6$  & 0.18 & 0.93 & 17 &  21 (15)  & 31 (20) & 38 (25) & 41 (26)  & 41 (29) & 67 (44)  \\
Ursa Minor  & $9.3 \pm 1.8$  & 0.20 & 0.64 & 16 &  21 (15)  & 31 (20) & 37 (25) & 41 (26)  & 42 (29) & 69 (43)  \\ 
Leo I       & $8.8 \pm 0.9$  & 0.22 & 0.82 & 15 &  17 (12)  & 24 (16) & 28 (19) & 30 (20)  & 33 (23)  & 51 (32)  \\
Carina      & $6.8 \pm 1.6$  &  0.21 & 0.69 & 12 &  11 (8)  & 14 (10) & 18 (13) & 18 (13)  & 22 (16) & 29 (19)  \\
Leo II      & $6.7 \pm 1.1$   & 0.16 & 0.48 & 12 &  13 (9)  & 19 (13) & 23 (16) & 24 (16)  & 27 (19) & 40 (26)  \\
Sculptor    & $6.6 \pm 0.7$  & 0.11 & 1.5 & 11 &  13 (9)  & 18 (12)  & 21 (15) & 23 (15) & 26 (18) & 38 (24)  \\
Sextans     & $6.6 \pm 0.7$  & 0.34 & 3.2 & 11 &  8 (7)   & 10 (7)  & 10 (8) & 11 (8)    & 12 (9) & 14 (10)  \\
\\
SMC         &                &    &    & 30 &  $< 60^*$  & $< 60^*$  &$< 60^*$  & $< 60^*$    & $< 60^*$ & $< 60^*$    \\
LMC         &                &    &    & 50 &   50     &   50    & 50     &  50       &  $50^*$  &  $50^*$  \\

\enddata

\tablecomments{The names  of the satellite  galaxies of the  Milky Way
are  given  in column  (1).   We  consider  only those  galaxies  with
galactocentric distances smaller than $300$ kpc.  Columns (2-4) give the
measured  line-of-sight velocity dispersion, and King core and tidal 
radii for each satellite.  The exceptions are the LMC and the SMC. These
galaxies have measured rotation speeds, listed in Columns (6-11).  All
velocities are expressed in  units of $\kms$ and distances are 
in kpc.  Column (5) gives the
value  of $\Vmax$  assigned to the halo of each satellite by
by  K99.
Columns (6-11) and above the horizontal space 
 give the  value of $\Vmax$  that we estimated
for  each satellite
based on its measured  velocity dispersion and King profile parameters
and assuming  the primordial  power spectra specified at the top of
each column.  Values  of $\Vmax$
listed  without  parentheses were  calculated  assuming an  anisotropy
parameter   of  $\beta=0$,   and  those   inside   parentheses  assume
$\beta=0.15$.  
Except for  the case of
Draco, we use the velocity dispersions and King profile core and tidal
radii  for the Milky   Way satellites given  in  the review article by
Mateo (1998).  For Draco, we use  the parameters quoted by Odenkirchen
et  al.  (2001).    The  quoted maximum rotation speeds for the LMC and
SMC were taken from van der Marel et al. (2002) and Stanimirovi{\'c} (2000).
The LMC rotation curve is observed to be flat from $4$kpc out to $>8.9$kpc.  
For many of the low-power models (indicated by superscript ``$^*$''), 
the flat portion of the curve ($\sim \Rmax$) is 
expected to be at larger radius.  In order to explain this in the
context of these models,
we must suppose that baryonic in-fall plays an important role in setting
the properties of dark matter  rotation curves (Blumenthal et al. 1986).  In this
case, the measured value of $\Vmax$  ($\Rmax$) is larger (smaller)
than it would be for the pristine halo prior to baryonic contraction.  
The SMC rotation curve is even
more likely to be influenced by baryons (Stanimirovi{\'c} 2000), and
baryonic in-fall is likely to be of some importance for all cases.  While 
not demanded by the data, the effects of baryonic in-fall
could be important for all satellites, thus the listed 
$\Vmax$ values should be considered {\em lower} limits.  Lastly, 
the large value of $r_{\rm max}$ associated with Draco in the 
$\sigma_8 = 0.65$ case may be difficult to reconcile with 
the kinematic data of Kleyna et al. (2002) and may disfavor a model 
with such low power.
} 
\end{deluxetable}

\twocolumn
%
%

In addition to the considerable uncertainties associated with galaxy formation, 
there are potential shortcomings in our efforts to model 
substructure properties in the context of collisionless dark matter physics 
alone.  For example, we have allowed for only a mild
redistribution of mass within the tidal radii of the orbiting subhalos
up until the time the subhalo is totally disrupted.  The work
of S02 and H03 suggests that this effect may be larger, but these
results may have been compromised by limited numerical resolution or
inappropriate assumptions regarding initial subhalo orbits and/or
accretion times.  In this sense, our approach represents a
conservative extreme because we assume that the surviving subhalo
density structure is typically very similar to that of halos in the
field.  We study the effects of tidal mass redistribution in a
forthcoming paper (J. Bullock, K. Johnston, \& A. Zentner, in preparation).  
In addition, we have adopted a  halo concentration relation (B01) that
has  not been confirmed for $M \lsim 10^9$ M$_{\odot}$.   Similarly, the EPS
merger tree calculations have yet to be tested in the very low-mass regime.
In light of these extrapolations, it is imperative that our results be
tested, and updated using the next generation of numerical studies.

Finally, our  model does not treat the substructure
population self-consistently.  We have neglected any subhalo-subhalo 
interactions which could serve to increase the internal heating of substructure and 
modify dynamical friction timescales as orbital energy is exchanged 
between subhalos and traded for internal energy.  We have adopted the approximation 
that all in-falling  halos are  ``distinct'' and have  no subhalos  of their
own (see Taylor \& Babul 2003 for a study of merger tree 
``pruning'').  However, our ``tree-level'' calculations suggest that the 
substructure mass fraction is uniformly $\sim  10\%$ regardless of 
host mass, so we expect  the that  this correction  would typically 
affect our derived mass fractions by $\lsim 10\%$.  Considering the 
assumptions that  have gone into  our calculations and the 
current level of observational precision, this level of error is 
acceptable; however, it may need to be improved upon as observations 
zero in on the masses of the subclumps responsible for the lensing 
signals and the mass fractions in these subclumps.

\section{\label{sec:conclusions}Conclusions and Discussion}

The abundance of substructure in dark matter halos is determined 
by a continuous competition between accretion and disruption.  
Accreted  subhalos with dense cores are resistant to disruption, but 
over time their orbits decay, their mass  is stripped away, and they are 
often destroyed.  The model we presented here allows us  to follow the 
complicated interplay between density structure, orbital evolution, and accretion time 
in order to determine how changes in the power spectrum affect the final
substructure population in  galaxy-sized  halos.  For  a  fixed set of
cosmological  parameters, changes in    the  power spectrum   manifest
themselves  by changing  collapse  times for halos,  where  less power
leads  to  later  accretion times    and  lower  densities.  We   have
specifically focused on tilted models that help to relieve the central
density crisis facing CDM and that may be favored by joint CMB and 
large-scale structure analyses (Spergel et al. 2003).  
We  have also considered a BSI inflation  model  and WDM
models, where the power is sharply reduced on small scales.

For a large class of CDM-type models, including models with significant
tilt and running, we find that the fraction of  mass  bound up in substructure 
$f$, for  galaxy-mass halos is relatively insensitive to the slope of 
the primordial power spectrum.  This is because both the host halos and their accreted
subsystems  collapse later in these models,  in a roughly self-similar
way as power  is reduced.  Note that this result would have  been roughly 
expected if we were varying  only the overall normalization in
the  models because  the  relative  redshifts  of   collapse would  be
invariant (we assume host halos are small enough that they collapse
before $z\sim  0$).  Our investigation suggests that   this   intuitive
description holds  even for tilted and  running index models, at least
over the  parameter range we  have explored.  All indications are that
this insensitivity to the tilt of the power spectrum is a rather robust 
result, and should hold even if some  unknown factor  has caused  our overall
normalization in predicted mass fractions to be in error (\eg our exclusion 
of central galaxies).  Interestingly, the shape   of the  mass  function, 
$f(x \equiv M_{\rm sat}/M_{\rm host})$,  is also relatively  insensitive to
the mass  of the host halo [Eq. (\ref{eq:fitfrac})], and a similar shape holds for all
of the tilted models we explored.\footnote{Note   that these self-similar
trends break down on cluster-mass scales since recent accretion, which
is a strong  function of the overall power,  likely plays an important
role in  these objects.}  

The similarity in mass  fractions breaks down for models
with sharp features in their power  spectra, like our BSI and WDM
models.  In these models, low-mass halo formation is delayed significantly
{\em relative} to the formation time of their hosts.  Consequently, fewer 
subhalos are dense enough to withstand the tidal field they experience
upon accretion.  We  find that for the relevant WDM and BSI models, 
the mass fraction in substructure is reduced by a factor of 
$\gsim 3$ compared  to the standard/tilted $\Lambda$CDM models.  

Inspired by recent attempts  to  measure substructure  mass  fractions
using multiply-imaged quasars,  we applied our  model to  ensembles of
host halos  with  $M = 3\times10^{12}$ \Msun at $z=0.6$, which 
represents the expectation for massive lens  galaxies  (DK01; DK02).  
For the $\Lambda$CDM/tilted cases, we found substructure
mass fractions  within  a $10$  kpc  projected radius in systems less
massive  than $M= 10^8$, $10^{9}$, and $10^{10}$\Msun of 
$f_8 \simeq 0.2-0.4\%$,   $f_9 \simeq 0.4 - 1.5\%$, and 
$f_{10} \simeq  0.6-2.5\%$ at the $64$ percentile range.  These
estimates are consistent with, but on the low  side of, first attempts
to measure the substructure fraction  using multiply-imaged quasars by
DK01, who obtain $f  \simeq 0.6\% - 7\%$ at $90\%$
confidence,  with an  upper-mass limit of $10^8-10^{10}$ M$_{\odot}$ (N.
Dalal, private communication).  The lensing results disfavor  the BSI
model, which leads to mass fractions $f_8 \simeq 0.01-0.06\%$, 
$f_{9} \simeq 0.02 -  0.2\%$, and $f_{10} \simeq 0.03 - 0.4\%$ at $64$\%.  
This is true unless the break scale in the power spectrum is pushed to 
such a  small value that this model no longer has the attractive feature 
of alleviating the central density problem.  A $m_{\rm W} =  0.75$ keV WDM
model  is similarly disfavored, and even our highest mass WDM case,
$m_{\rm W} = 3$ keV, has a typical projected fraction ($f_9 \sim
0.4\%$) that is low compared to the DK01 estimate.  
Again, this indicates that if  the warm particle is a thermal relic, 
the mass must be large  enough that WDM no longer mitigates the 
small-scale problems of standard CDM.  Yet, these
results are interesting  because they show how  lensing may be
used as one of the few probes of the WDM particle mass in the
range  $\gsim 1$ keV or a  break in  the  primordial power spectrum at 
large wavenumber.

Clearly these conclusions  must be regarded with some caution.  In
addition to the  uncertainties of modeling discussed in \S \ref{sec:caveats}, 
other issues make drawing definite conclusions difficult.  
For example, we have only  accounted for the substructure
within the virial radius of the   host halo, yet  the anomalous flux
ratios of lensed images are sensitive to the presence of small halos
along the line-of-sight to the lens.  Keeton (2003) showed that field
halos can have a significant lensing effect even if they are separated
from the lens by several tenths in redshift and  in  hierarchical,
CDM-type models, small field halos are ubiquitous.  Chen et  al.  (2003) 
showed that the relative effect from halos outside the virial radius of 
the lens is  typically a few percent, but may be as large as 
$20 - 30\%$ of that from subhalos,  depending upon assumptions 
about the subhalo population.  Also, as the  mass
fraction  in substructure of a given mass depends on  the mass of the
host [Eq. (\ref{eq:fitfrac})], it may be important to constrain the host halo mass in
order to fully exploit the ability of lensing  measurements of
substructure to probe cosmology and structure formation.

We compared our model predictions for the cumulative subhalo velocity
function, $N(>\Vmax)$, to the satellite galaxy count of the Milky Way.
The approach here was to estimate $\Vmax$ for each satellite
galaxy's dark matter halo based on its observed line-of-sight velocity
dispersion,  $\svd$.  We emphasized in \S \ref{sub:dwarf} that the
mapping  between   $\svd$ and $\Vmax$  is sensitive to {\em
theoretical} prejudice regarding the density  structure of the dwarf 
galaxy's  halo as well as the {\em unknown}  velocity anisotropy 
parameter of  the  system, $\beta$.  
For a  fixed value of $\beta$,  less concentrated host halos
imply {\em larger} values of $\Vmax$ because  halo rotation curves are
more slowly rising and stars probe only the  inner $\sim 1$ kpc of the
halo.  Interestingly, this implies that tilted models  and truncated
models, do  significantly better than $n=1$, $\Lambda$CDM in reproducing
apparent dwarf counts, even though their mass fractions are similar.
While our estimates of $\Vmax$ cannot be considered robust because of the 
simplicity of our model and the fact that the $\svd-\Vmax$ mapping is 
very sensitive to the inner structure of the subhalos, the general trends 
that we illustrate should persist in more elaborate studies and 
N-body simulations.  Moreover, our results reveal yet another reason why 
it is difficult to consider the dwarf satellite problem a serious 
challenge to CDM theory, the nature of the dwarf satellite problem 
is very sensitive to cosmology, the power spectrum, and assumptions about 
the shape of the velocity ellipsoid for stars in dwarf galaxies.

When we fix  $\beta=0$, our ${\rm d}n/\rm{d}\ln k  = -0.03$ RI model, 
$\sigma_8 = 0.75$ model, and the BSI case all do  well in matching the
known satellite population of the Milky Way for $\Vmax \gsim 20 \kms$.
Our lowest power model ($\sigma_8  = 0.65$, $n  \simeq 0.84$, and mild
running) actually under-predicts the  dwarf count for $\Vmax \gsim$ 30$
\kms$.  However, this result   is  achieved only for  the  optimistic
assumption of isotropic velocities.  If  we  adopt a small level
of anisotropy, $\beta = 0.15$, consistent with the centers of simulated 
dark matter halos, agreement for most models is worsened.  
Only the BSI and $\sigma_8 = 0.65$ models show good
agreement in this case.  Yet, even with $\beta = 0.15$, the RI 
and $\sigma_8 =  0.75$ models still compare
more favorably than the $n=1$ case with $\beta=0$.

What do these  results imply for  the dwarf satellite problem?  In all
models, including those with truncated power, the velocity function of
subhalos  continues to rise below  the  scale of the smallest observed
Milky Way satellite, $\Vmax \lsim 10 \kms$.  No matter how
one modifies the power spectrum, some kind  of feedback is required to
explain the local satellite population.  Different power spectra 
(even different values of $\beta$) seem to indicate that  different
types of  feedback are  needed.  For  example,  in models  with $\sig8 \gsim 0.8$ 
(the precise number depends on typical $\beta$ values and
the degree of running/tilt), the feedback must be {\em differential}.
That is, for $\Vmax \simeq  8 - 30 \kms$, only  one out of every $\sim
5-10$ halos in this range should form stars.  On the other hand,
in  models like the ${\rm d}n/{\rm d}\ln k = -0.03$ RI model, 
the BSI case, and our $\sig8 \simeq 0.75$ 
model with $\beta=0$, the discrepancy  seems to set in suddenly
at $\Vmax \sim 10-20 \kms$, suggesting that nearly all halos smaller
than this are completely devoid of stars.  In this case, the feedback
mechanism must provide a sharp transition.

The feedback mechanism proposed by BKW accommodates the need for only 
$\sim 10\%$ of subsystems to actually host observable galaxies 
by suggesting that only those systems that formed 
before reionization  were  able  to  retain   their  gas  and
eventually form stars.  However, if  reionization were  to occur very
early (\eg $z \gsim 15$), many fewer  than  $10\%$  of  these
dwarf-sized systems could have  collapsed before reionization, so that
almost  all systems smaller  than $\Vmax \sim 30  \kms$ would be dark.
This would be more in line with what we  see for the low-power models.  
This is an intriguing result.  The best-fit power 
spectrum of the WMAP team (Spergel et al. 2003) 
leads to similar substructure mass fractions as standard CDM, 
alleviates the central density problem and dwarf satellite discrepancy, 
and forces us to consider feedback mechanisms that predict a sharp 
transition between luminous and non-luminous galaxies. Additionally, 
the possible detection of early reionization by the WMAP team (Kogut et al. 2003) 
{\em provides} a feedback mechanism that results in a sharp transition.  
Of course, explaining early reionization in models with low
small-scale power may be problematic (Somerville et al. 2003).  
Another feedback scenario that leads to a sharp transition is 
photo-evaporation (Barkana \& Loeb 1999; Shaviv \& Dekel 2003). 
Nonetheless,  the uncertainty associated with $\beta$ in determining 
satellite galaxy $\Vmax$ values suggests that efforts to model dwarf 
galaxy  luminosities as well as dynamical  properties  will be  
required to resolve this issue (Somerville 2002; Benson et al. 2002).

Remaining uncertainties in understanding the precise nature 
of the dwarf satellite problem highlight the need to focus on 
attempts to measure CDM substructure by other means.  Continued 
efforts to detect substructure via gravitational lensing 
(\eg DK01; Keeton 2003; Keeton, Gaudi, \& Petters 2002; 
Moustakas \& Metcalf 2003) or by probes within our own Galaxy 
(\eg Johnston et al. 2002; Ibata et al. 2002a, 2002b; 
Mayer et al. 2002; Font et al. 2001) offer useful avenues 
for doing so.  Modeling of the kind presented here may play an important 
role in interpreting results of ongoing observational programs and 
bring us one step closer to confirming or refuting one of the fundamental 
predictions of the CDM paradigm.
%
%

\begin{acknowledgments}

We are pleased to thank Neal Dalal, Kathryn Johnston, Charles Keeton,
James Kneller, Chris Kochanek, Leon Koopmans, Andrey Kravtsov, 
Eugene Lim, Ari Maller, Benton Metcalf, Leonidas Moustakas, Joel Primack, 
David Rusin, Bob Scherrer, Rachel Somerville, Gary Steigman, Rob Swaters,  Terry
Walker, Risa Wechsler, and David  Weinberg for many helpful discussions.  
We are also grateful to Neal Dalal, Michael Schirber,
Risa Wechsler, David Weinberg, and Joshua Winn for valuable comments on
an early draft of this manuscript.  JSB acknowledges Rosemary Wyse for
her  generous hospitality during visits   to Johns Hopkins University,
where a substantial fraction of this  work was done.  ARZ is supported
by      The Ohio   State University      and  by   U.S. DOE   Contract
No. DE-FG02-91ER40690.   JSB  is   supported by NASA   through  Hubble
Fellowship grant    HF-01146.01-A  from the  Space   Telescope Science
Institute, which  is operated by  the Association  of Universities for
Research in Astronomy, Incorporated, under NASA contract NAS5-26555.

\end{acknowledgments}



\end{document}